\documentclass[lineno]{jfm}
\usepackage{subcaption}
\usepackage{graphicx}
\usepackage{newtxtext}
\usepackage{newtxmath}
\usepackage{natbib}
\usepackage{hyperref}
\usepackage[dvipsnames]{xcolor}
\hypersetup{
    colorlinks = true,
    urlcolor   = blue,
    citecolor  = MidnightBlue,
}

\newcommand{\RomanNumeralCaps}[1]
\linenumbers

\usepackage{silence}
\title{Airfoil tonal noise reduction by roughness elements \\ Part II -- Direct simulations}

\author{Zhenyang Yuan\aff{1}
  \corresp{\email{zhenyang@kth.se}},
  El\'ias Alva\aff{2},
  Tiago B. de Ara\'ujo\aff{2},
  Andr\'e V. G. Cavalieri\aff{2},
 \and Ardeshir Hanifi\aff{1}}

\affiliation{\aff{1}FLOW, Department of Engineering Mechanics, KTH Royal Institute of Technology, Stockholm, Sweden
\aff{2}Divis\~ao de Engenharia Aeron\'autica, Instituto Tecnol\'ogico de Aeron\'autica, S\~ao Jose dos Campus, Brazil}

\begin{document}

\maketitle

\begin{abstract}
In a combined experimental and numerical effort we investigate airfoil tonal noise generation and reduction. The means of noise control are streak generators in form of cylindrical roughness elements. These elements are placed periodically along the span of airfoil at the mid chord streamwise position. Experiments are performed for a wide range of Reynolds number and angle of attack in a companion work \citep{Alva2024arxiv}. In the present work we concentrate on numerical investigations for a further investigation of selected cases. We have performed wall-resolved large-eddy simulations for a NACA 0012 airfoil at zero angle of attack and Mach 0.3. Two Reynolds numbers  ($\mathbf{0.8\times 10^{5}}$ and $\mathbf{1.0 \times 10^{5}}$) have been investigated, showing acoustic results consistent with experiments at the same Reynolds but lower Mach numbers. Roughness elements attenuate tones in the acoustic field and, for the higher Reynolds number, suppress them. Through Fourier decomposition and spectral POD analysis of streamwise velocity data, dominating structures have been identified. Further, the coupling between structures generated by surface roughness and instability modes (Kelvin-Helmholtz) of shear layer has been identified through stability analysis, suggesting stabilisation mechanisms by which the sound generation by the airfoil is reduced by the roughness elements.
\end{abstract}

\section{Introduction}

Trailing edge noise, often referred to as airfoil self-noise, is a key factor contributing to noise pollution particularly relevant in aviation and wind energy applications. This issue is especially pertinent for communities situated near airports or wind farms. Due to its considerable impact, there is a focused effort to thoroughly understand, accurately model, and effectively mitigate trailing edge noise. Such initiatives aim to develop technologies that are not only quieter but also more eco-friendly. Trailing edge noise is generated due to the interaction between an airfoil blade and the turbulence produced in its own boundary layer and near wake \citep{BrooksThomasF.1989Asap} and is one of the main sources of noise pollution for airfoils moving through non-turbulent flows. 

Tonal noise, often induced by airfoils at low to moderate Reynolds number \citep{ArbeyH.1983Ngba}, is well studied in the aeroacoustic community. Compared with the broadband noise, tonal noise always shows as sharp peaks in the acoustic spectrum. Those peaks are always several dBs higher than the broadband noise and, thus, more displeasing and harmful to human ears. An improved understanding of the noise generation mechanism can lead to effective control strategies for noise elimination or abatement.

In the early research on the tonal noise produced by airfoils, \cite{paterson1973vortex} identified ladder like structures where the spectrum contains the main tone (also denoted as the primary tone \cite{arcondoulis2010review}) and secondary tones by the side of the main tone. \citet{TamChristopherK.W.1974Dtoi} suggested that the ladder-like structure is due to a self-excited feedback loop between disturbances in the boundary layer and the airfoil wake. 
In the subsequent work, \citet{ArbeyH.1983Ngba} explained the phenomenon as the result of a feedback mechanism involving instability waves in the boundary layer and acoustic waves that are scattered at the trailing edge and travel upstream, exciting the instability waves and thus closing the loop. Linear analysis by \cite{FosasdePandoMiguel2014Agao} and \cite{RicciardiTulioR.2022Tiap} confirms the said mechanism.

The regime of the tonal noise on an airfoil with laminar and turbulent flows has been studied by \citet{ProbstingS.2015Rotn}. The study shows that regimes of tonal noise generation can be separated into pressure- and suction-side dominated regions. This provides a condition for  amplified spanwise coherent vortical structures at the pressure or suction side of the airfoil. Such structures will eventually pass by trailing edge which induce or contribute to acoustic tonal noise radiation. Experiments with forced transition to turbulence shows that lower Reynolds number tonal noise emission is dominated by suction side while higher Reynolds number emission is dominated by pressure side.

To control tonal noise generation, a promising idea is to attenuate spanwise coherent structures, as tonal noise in airfoils is mostly related to 2D disturbances \citep{RicciardiTulioR.2022Tiap}. In the work of \cite{FranssonJensH.M.2005Esot}, experiments shows that stable and symmetric, close to sinusoidal, streaks of moderate amplitudes (12\% of the free-stream velocity) can stabilize Tollmien-Schlichting waves (T-S waves). The streaks are generated by the means of a spanwise array of cylindrical roughness elements. In the study, the stabilizing role of the streaks on TS waves is unambiguously confirmed by increasing the height of the roughness elements. 
Numerical simulation (\cite{CossuCarlo2002SoTw}) and linear stability analysis (\cite{CossuCarlo2004OTwi, Bagheri2007TheSE}) also confirm that stabilization effect of streaks on T-S waves at least up to a local Reynolds number ($R$) of 1000 which $R$ is given by $R = \sqrt{XRe_L}$, where $Re_L$ is a constant and $X$ is the streamwise location. Furthermore, this stabilization is due to the spanwise averaged part of the nonlinear base flow distortion induced by the streaks and occurs for streak amplitudes lower than the critical threshold beyond which secondary inflectional instability is observed. This modulated T-S wave has almost identical phase speed but lower growth rate than the corresponding two dimensional T-S waves.

In the work of \citet{MarantMathieu2018Iooa}, it also shows that streaks can stabilize Kelvin-Helmholtz (K-H) instabilities. When forced with finite amplitudes, the streaks modify the characteristics of the Kelvin–Helmholtz instability. Maximum temporal growth rates are reduced by optimal sinuous perturbations and are slightly increased by varicose sub-optimal ones. In contrast, the onset of absolute instability is delayed by varicose sub-optimal perturbations and is slightly promoted by sinuous optimal ones. Besides modifying Kelvin-Helmholtz instabilities, streaks can render separation bubbles three-dimensional, as explored by \cite{KarpMichael2020Osoa}.

Direct numerical simulations of the trailing edge noise is a relatively new research topic. \citet{WangMeng2000CoTF, SINGERB.A.2000SOAS, ManohaEric2000TNPU} are among the first to use imcompressible large eddy simulations (LES) to compute near field unsteady flow around the blades or thick plates and use acoustic analogy of Ffowcs Williams and Hall \citep{WilliamsJ.E.Ffowcs1970Asgb} to predict the far-field acoustic sound. More recently, other acoustic analogies, i.e. Curle's \citep{CurleN.1955Tios} and Ffowcs Williams and Hawkings (FWH) acoustic analogy formulations have been used for far-field acoustic prediction. With increasing computational power and more modern computational fluid dynamic codes, direct simulation of far-field acoustics becomes possible. \cite{Sandberg_aiaa_14th, SandbergR.D.2013DNSf, Jones_aiaa_2010} conducted direct numerical simulation (DNS) of transitional/turbulent flow at moderate Reynolds number past NACA 0012 airfoils at different angles of attack. They investigated mechanisms of noise  generation, attempting to identify sources of airfoil noise other than trailing edge noise.

Inspired by the results stated above, we investigate the tonal noise generation of aerofoils and its control through experimental and numerical studies. The control means is a row of spanwise periodically placed cylindrical roughness elements. These roughness elements, if not too large, generate laminar streaky structures which are expected to cause modulation and weakening of two-dimensional structures over the wing. In the first part of the current work reported in the companion paper \citep{Alva2024arxiv}, a series of experiments with different Reynolds number and angles of attack have been conducted for a NACA0012 airfoil. In these experiments, both clean aerofoils and the ones with periodically placed roughness elements have been investigated. Critical roughness Reynolds  number ($Re_{kk} = U_k$k$\rho_k/\mu_k$, where k is roughness height and subscript $k$ represents quantities at roughness height) based on roughness height has been controlled to be $Re_{kk} = 366$ to make sure boundary layer will not be tripped directly via roughness for the case with chord based $Re = 100,000$ in the simulation. According to calculation from \citet{FranssonJensH.M.2005Esot} on a flat-plate boundary layer, our $Re_{kk}$ is far less than critical $Re_{kk} = 420$. At low Reynolds number, i.e. $Re = 80,000$, both geometries can generate tonal noise. Lower amplitude and different frequency of tonal noise are detected in the setup with roughness elements. At slightly higher Reynolds number, i.e. $Re = 100,000$, tonal noise has been almost eliminated by roughness elements. Similar to \citet{ProbstingS.2015Rotn}, at the different angles of attack, trailing edge noise is dominated by either pressure side or suction side coherent structures, which can be inferred by studying separately the effects of pressure- and suction-side roughness on sound radiation.  

In the present work, we will try to investigate the mechanism of  tonal noise attenuation or elimination in the presence of streaks generated by roughness elements. A series of wall-resolved large eddy simulations (LES) are performed to study the problem. The problem setup analyzed here is the same as in the experiments, namely a NACA 0012 airfoil at zero angle of incidence  and chord-based Reynolds number $Re = 80,000$ and $100,000$. Note that due to slow convergence of numerical simulations at very low Mach numbers, the simulations are performed at a higher Mach number of 0.3 compared to the experiments. Both of the airfoil with/without the surface roughness elements are investigated and compared. Four cases in the following context are referred as Re = 80k clean, Re = 80k rough, Re = 100k clean and Re = 100k rough, respectively.  Details of the numerical simulation tools and methodology of analysis are presented in the \S \ref{sec:NUM}. \S \ref{sec:FD} contains the results from fluid dynamics and acoustic measurements. Finally, the data driven spectral proper orthogonal decomposition (SPOD) and stability analysis at selected streamwise positions are carried out in order to understand the influence of structures generated by the roughness elements on the flow. These analyses are presented in the \S \ref{sec:SPODST}. The paper is completed with conclusions in \S \ref{sec:conclusions}.
\section{Numerical setups}
\label{sec:NUM}

\subsection{Large eddy simulation}
Prior to the simulation, a series of experimental campaigns have been performed. In order to be consistent with the experimental setup, the airfoil profile used in the simulations will be the same as in the experiment. Readers are encouraged to read the companion work for more information about the experimental setups \citep{Alva2024arxiv}. Here, a short summary on geometry and flow conditions will be introduced.

The geometry investigated here is a modified NACA0012 airfoil. The airfoil has zero angle of attack and the free-stream Mach number is set to be $M = U_{\infty}/a_{\infty} = 0.3$, where $U_{\infty}$ is the freestream velocity and $a$ is the freestream speed of sound. Two values of Reynolds number, $0.8\times 10^{5}$ and $1.0 \times 10^{5}$, have been chosen to be investigated for potential different flow phenomena. The chord length of the model is $c$ = 100 mm. Unlike the theoretical NACA0012 profile, the trailing edge is truncated at around 98\% of the chord length and rounded with an arc-circle of radius $0.04c$, in order to control the trailing edge shape, in agreement with earlier works of \citet{RicciardiTulioR.2022Tiap, RicciardiTulioR2022Sotn}. Boundary-layer displacement thickness at the trailing edge location is expected to be much thicker than the trailing edge thickness. Thus trailing edge bluntness is not expected to be dominant for trailing edge noise \citep{BrooksThomasF.1989Asap}. 

We consider two different geometries in the present study: one with smooth surface and one with a row of spanwise-periodic roughness elements on both sides of airfoil. The choice of roughness elements follows the work by \citet{FranssonJensH.M.2005Esot}. Roughness elements have height of $0.0055c$ and diameter of $0.015c$. The distance between them is $0.06c$. These roughness elements are placed close to the mid-chord position, at $x/c=0.52$. The extension of the computational domain in the spanwise direction is $0.12c$, which contains two periods of roughness elements. A periodic boundary condition is used in the spanwise direction.  According to the scattering condition, in order to generate propagating acoustic waves, spanwise wavenumber has to satisfy $k_z < k_0$ where $k_0$ is the acoustic wavenumber \citep{Nogueira2017Ampf, Sano2019, demange2024, Yuan2024_wavepacket}. In this study, the spanwise coherent two-dimensional flow structures ($k_z = 0$) dominate the generation of tonal noise. Thus the choice of spanwise width is expected to have a small effect on the noise generation.

The open source high-order Flux-Reconstruction numerical framework PyFR \citep{WITHERDEN20143028} is used for wall-resolved implicit large eddy simulations (iLES) to solve the compressible Navier-Stokes equations in the general Cartesian coordinates with geometries stated above. Unlike LES with sub-grid scale models, iLES introduces the numerical dissipation by the discretization. In this simulation, we use anti-aliasing technique via approximate $L^2$ projection, which proved to be numerically stable and efficient. The principle behind this anti-aliasing is to compute the modal expansion coefficients of the desired polynomial exactly, see \citet{ParkJ.S2017HILS} for details.
The discrete set of equations obtained from the spatial-spectral (hp) type discretization are integrated in time using a TVD-third-order Runge-Kutta method. The simulations are performed using polynomial order 4. 

PyFR supports both structured and unstructured mesh topology. To benefit from this, we used meshing strategy as following: wall resolved O-grid with maximum resolution at around mid-chord is given in terms of wall unit: $\Delta x^+ < 20$, $\Delta z^+ < 10$, $\Delta y^+ < 0.9$, where the flow is basically laminar. Considering the flow on the airfoil is still transitional or fully developed to turbulence in the high Re cases, thus at the trailing edge, resolution in streamwise and wall normal direction is kept as  $\Delta x^+ < 4$, $\Delta y^+ < 0.7$. Near wall mesh elements are lifted to second order to fit geometry curvatures. The structured mesh is also applied in the wake resolved region until $1.5c$ after the trailing edge. The first layer of elements from the airfoil surface has initial height
$\Delta s = 0.00055c$ and growth of the structured layers has the rate of 1.08.
For the cases with roughness element, a prism type of mesh are applied in the near roughness region to have the better fit of roughness shapes and avoid stretch of the grid. The prism layer has much smaller elements with $\Delta s = 0.002c$ which can reduce mesh-to-mesh discontinuity due to the discontinuous nature of Flux Reconstruction method. 

In this work, length scales, velocity components, density, pressure and frequency are non-dimensionalized as $x = x^* /c $, $u = u^* /a_{\infty}$, $\rho = \rho^* /\rho_{\infty}$, $p = p^*/(\gamma a_{\infty}^2 \rho_{\infty})$ and $St = f^*c^*/U^*_{\infty}$ respectively. Here, $\rho_{\infty}$ is freestream density, $gamma$ is ratio of specific heats and the quantities with superscript $*$ are given in dimensional units. Therefore, frequency $St$ presented here is a Strouhal number. Furthermore, the Sutherland’s law is applied for viscosity correction.

\begin{figure}
    \centering
    \includegraphics[width = 0.8\linewidth]{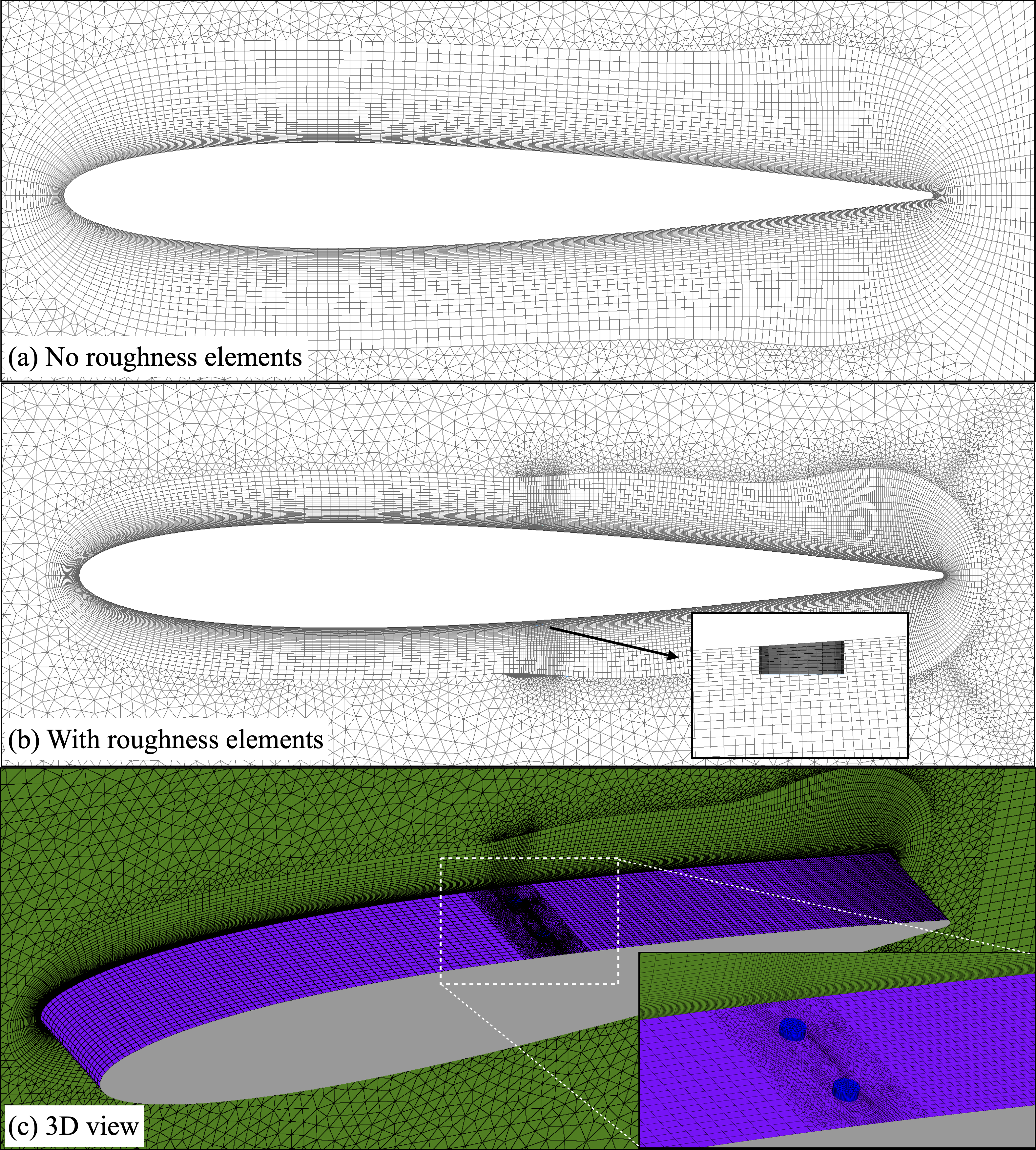}
\caption{High order computational mesh elements (polynomial order of 4) near the airfoil, a): without roughness elements (clean case); b): with roughness elements (rough case); c): 3D view of the wall surface mesh with roughness elements. Grid points inside the elements are skipped for a better visualization. Structure grids are applied in the near wall region while unstructured grids are applied in acoustic region.}
    \label{fig:mesh}
\end{figure}

In the acoustic region, unstructured grids are applied to reduce large aspect ratio and significantly reduce computational size. Analysis of the simulations data shows that acoustic waves are still resolved up to Strouhal number $St = 15$. The element aspect ratio in this region is kept below 5. Meshing details are showed in the figure \ref{fig:mesh}. The total grid number is around 70 million (polynomial order of 4). 
To allow a proper representation of the acoustic field and have less interference from boundary conditions to the airfoil, the computational domain has been extended to 15 chords  away from the airfoil surface. A designed sponge zone is added around the outer boundaries to absorb hydrodynamic wake and acoustic waves \citep{FreundJ.B1997PIBC, BodonyDanielJ.2006Aosz}. 
The sponge is added as a forcing term $-\sigma(q-q_0)$ to the solver. Here, $q_0$ denotes freestream quantities in the conservative format and the sponge strength parameter $\sigma$ is carefully chosen such that no wave reflection is allowed. At the outflow boundaries we apply the non-reflective character-Riemann-invariant boundary conditions, which, together with the sponge zones, minimise reflections at the boundaries. At the inflow boundary also the character-Riemann-invariant condition is used to provide homogeneous potential flow with $M$ = 0.3. A periodic boundary condition is used in the spanwise direction. A non-slip-adiabatic-wall condition is applied to airfoil surface.

All simulations are performed on LUMI-G AMD MI250X 128Gb GPU nodes. A compressible Navier-stokes solver of $4th$ order polynomial with hip-backend is used. The quadrature order used in the simulation is one order higher than solver's polynomial order to fit requirement of iLES \citep{ParkJ.S2017HILS}. Interior solution
points (i.e. Gauss-Legendre points for the Hexahedral elements) are used for anti-aliasing purpose. At least 2800 snapshots (around 84 flow overs) are taken for all cases for spectral analysis. At least 30 flow overs are discarded before collection of snapshots. Around 4000 GPU hours are consumed for each case.

\subsection{Mesh validation}
The mesh of the current simulation is designed targeting at the higher Reynolds number, i.e. Re = 100,000. The mesh has distribution of points in the streamwise and spanwise directions on the airfoil surface given by $n_x = 830$ and $n_z = 130$, respectively. From the airfoil surface to the height of roughness height ($\triangle y = 0.0055c$), points distribution in wall normal direction is given as $n_y = 45$. For clarification, we name this mesh as $M0$ which has total 4.86 million grid points in this near wall region. A refined mesh $M1$ is designed as $n_x = 1185$, $n_y = 50$ and $n_z = 170$ which leads to 10.07 million total grid points. The refinement is more concentrated from roughness elements to the trailing edge region where the flow gradients introduced by flow transition are much larger. Moreover, the aspect ratio is maintained. 

\begin{figure}
    \centering
    \begin{subfigure}{0.45\textwidth}
    \centering
    \includegraphics[ width=\linewidth]{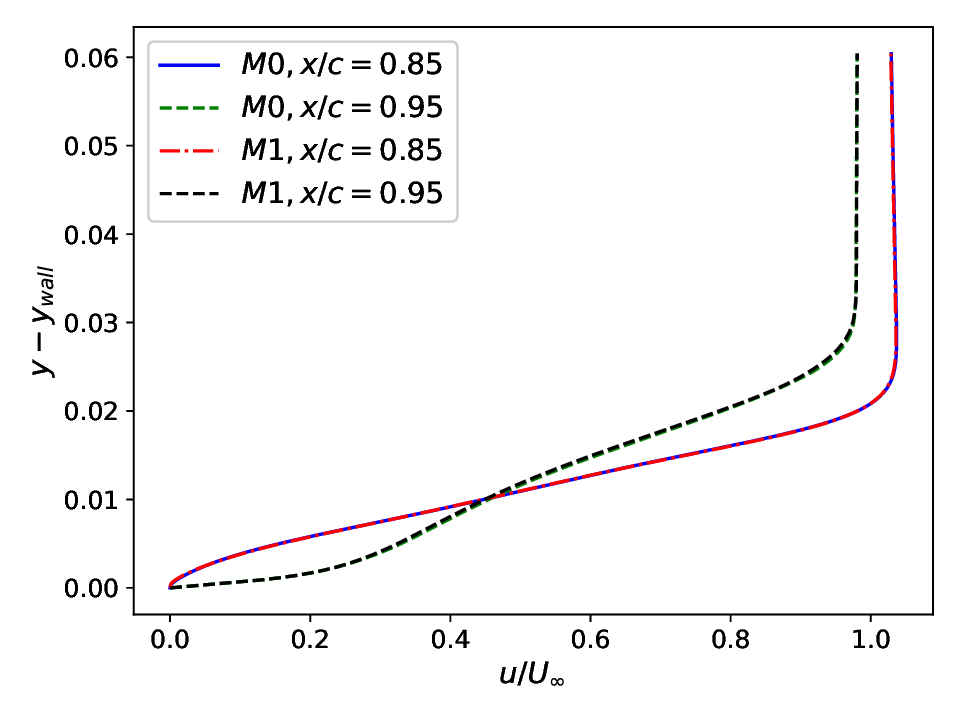}
    \caption{}
    \end{subfigure}
    \begin{subfigure}{0.45\textwidth}
    \centering
    \includegraphics[ width=\linewidth]{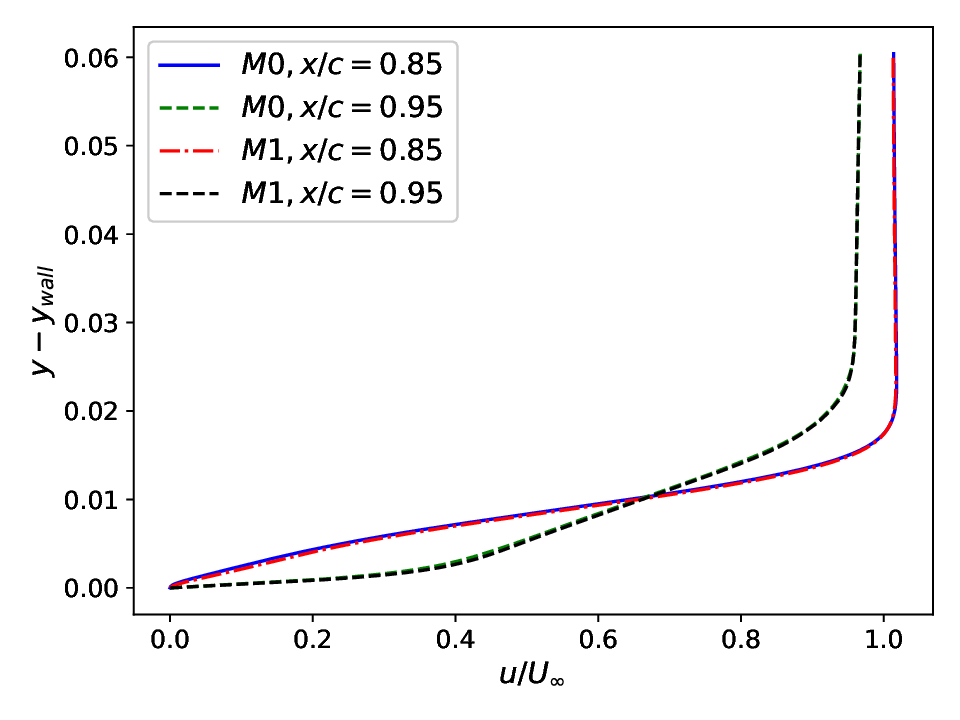}
    \caption{}
    \end{subfigure}
    \begin{subfigure}{\textwidth}
    \centering
    \includegraphics[ width=0.45\linewidth]{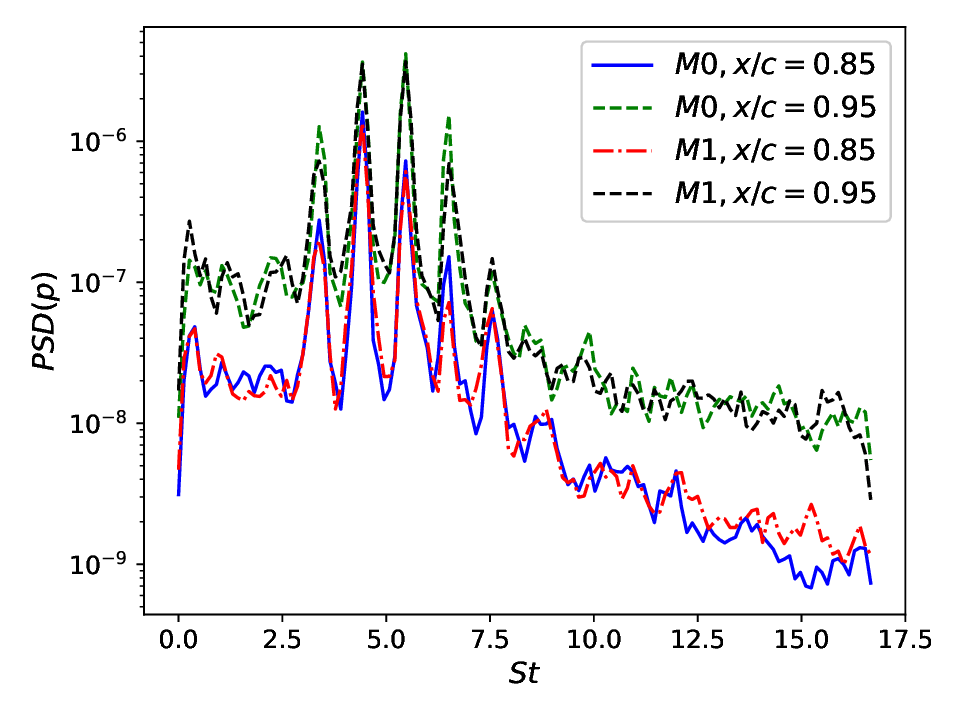}
    \caption{}
    \end{subfigure}
\caption{Mesh refinement study. Mean flow profiles measured for roughness cases at $x/c = 0.85$ and $x/c = 0.95$ with (a) Re = 80k and (b) Re = 100k, respectively; (c): acoustic spectrum measured at $x/c = 1$ and $y/c = 1$ for Re = 80k case.}
    \label{fig:mesh_independent}
\end{figure}

Figure \ref{fig:mesh_independent} illustrates the comparison between two mesh designs. Velocity profiles are acquired at streamwise location $x/c = 0.85$ and $x/c = 0.95$. Velocity profiles show great convergence as overlapping for both mesh setups. The acoustic spectrum measured at the case Re = 80k rough is shown for the location $x/c = 1$ and $y/c = 1$. Both mesh setups can capture the tonal peaks and secondary tones. There are slight differences due to the fact that the time series acquired for the $M1$ mesh is shorter than $M0$. For the case without roughness elements, the flow around the airfoil is expected to display weaker gradients, such that our mesh resolution $M0$ is more than sufficient. Note that both mesh setups use $4th$ order polynomial and $h$-type refinement is carried out. Since our cases are performed with implicit LES with anti-aliasing via projection acting on surface flux \citep{ParkJ.S2017HILS}, $p$-type refinement, i.e. increasing the polynomial order, will lead to the change of the filter as well as changing the effect of the filter caused by the time scheme. Thus, $p$-type refinement with the LES filter for grid convergence study is not considered in this study.

\section{Simulation results}
\label{sec:FD}
\subsection{Acoustics}
\begin{figure}
    \centering
    \begin{subfigure}{\textwidth}
    \centering
        \includegraphics[width = 0.8\linewidth]{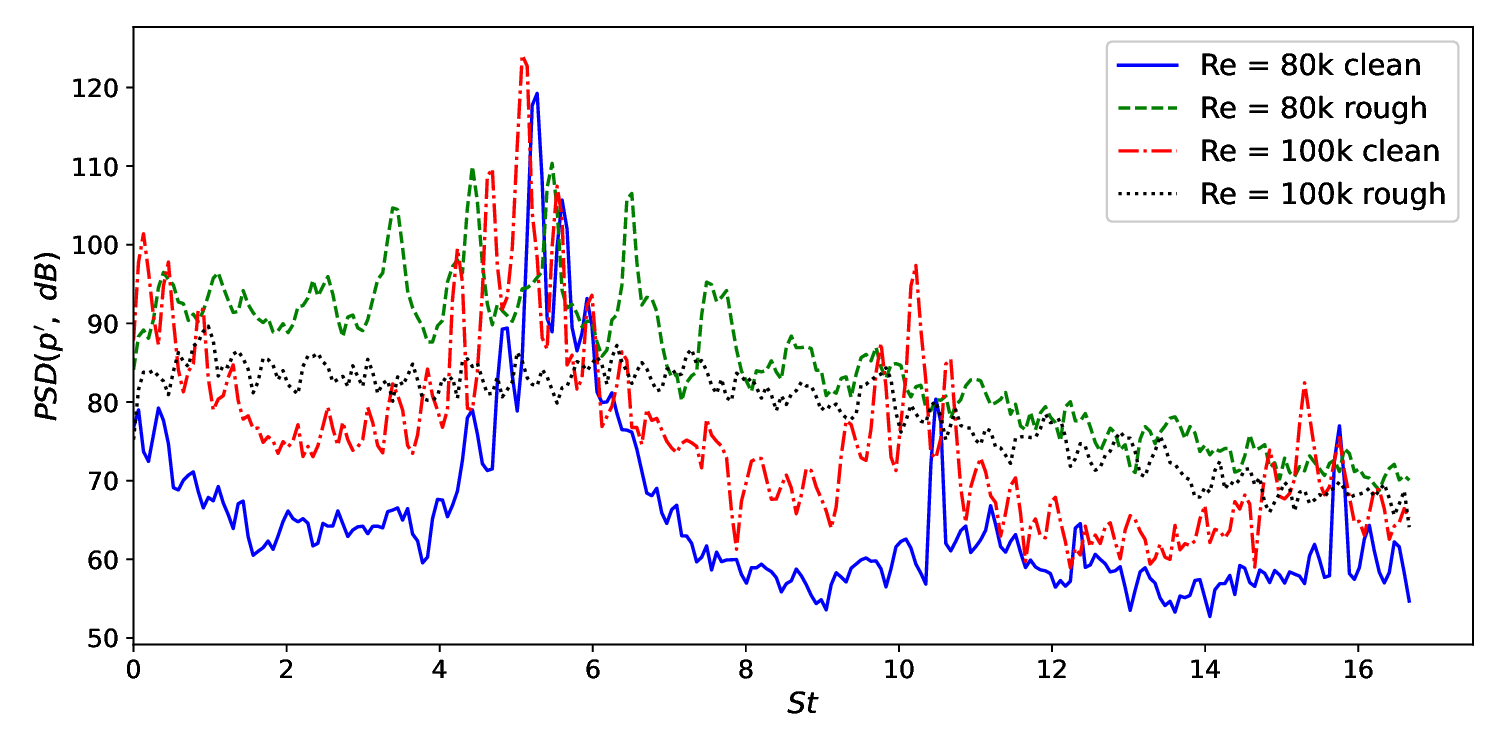}
        \caption{Numerical results}
    \end{subfigure}
    \begin{subfigure}{\textwidth}
    \centering
        \includegraphics[width = 0.8\linewidth]{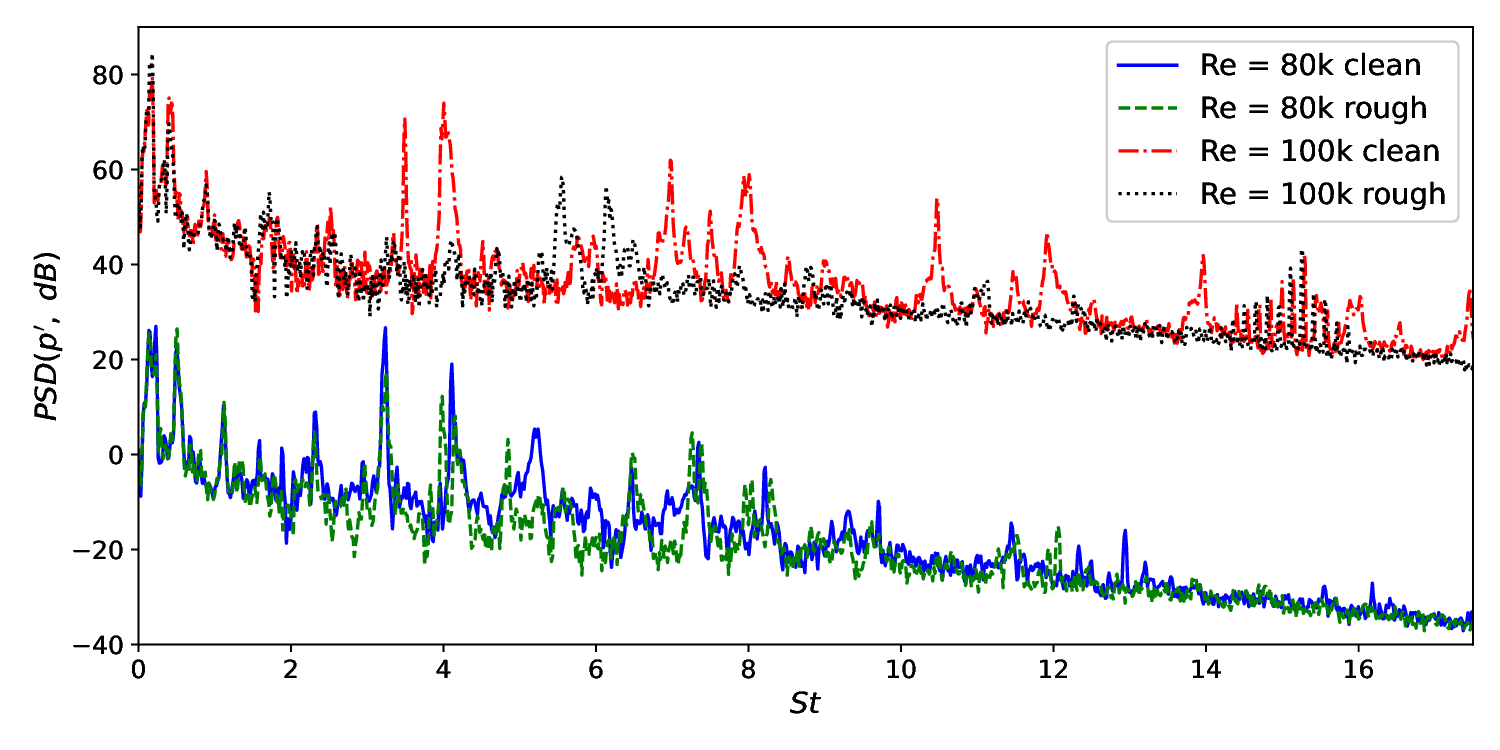}
        \caption{Experimental results. Results of Re = 80k cases are shifted 50 $dB$s lower for better visualization.}
        \label{exp_acous_spectrum}
    \end{subfigure}
\caption{Acoustic spectrum for numerical simulations and experiments measured in $dB$.}
    \label{fig:spectrum}
\end{figure}

In order to compare with the experimental results, the power spectrum density (PSD) of pressure fluctuations at the location $x/c = y/c = 1$ is illustrated in figure \ref{fig:spectrum}. The origin is taken at the leading edge of the airfoil, and this location is thus at 90 degrees from the trailing edge. In the experiments, a beamforming technique is applied to calculate the PSD of pressure fluctuation, and a detailed description can be seen in the first part of this work. A Welch method (\cite{WelchP.1967Tuof}) is used to calculate the PSD. For the LES data, the segment length  of Fourier transform is 512 and overlap percentage is set to 75\% which results into 22 blocks. The length of the time signal is 90 flow overs ($c/U_{\infty}$) with sampling frequency $St = 33.3$ to ensure a well converged dataset. Results are presented here in terms of the sound pressure level, which is defined in decibel ($dB$) as
\begin{equation}
    L_p = 10\log_{10}\frac{P_{xx}}{p^2_0}
\end{equation}
with a reference sound pressure $p_0 = 2 \times 10^{-5}$ $Pa$; we denote the power spectrum density of pressure fluctuations by $P_{xx}$. 

For all clean cases, the spectrum is dominated by a main frequency of $St = 5.47$ in Re = 80k clean case and $St = 5.21$ in Re = 100k clean case. With increasing Reynolds number, the main tone shifts to slightly lower frequency but it has stronger tonal noise amplitude. It is evident that secondary tones emerge in the spectrum at frequencies slightly lower and higher than the primary tone. As suggested by \citet{TamChristopherK.W.1974Dtoi}, the combination of the primary tone and the secondary tones is also referred as ladder like structures.
With roughness elements being added to airfoil surface, the main tone in Re = 80k case has been broken down into multiple tones with much lower amplitude and wider frequency bands. Those tones have more or less similar amplitudes, and it becomes harder to distinguish a primary tone in the spectrum. In the Re = 100k case, the spectrum of the rough case becomes of broadband nature. Both the main tone and secondary tones are eliminated.

Figure \ref{exp_acous_spectrum} presents the experimental measurement of the acoustic spectrum. A direct comparison between simulation and experiment is not possible, as will be discussed later, but similar trends appear. For visualization purposes, results for Re = 80k cases are shifted 50 $dB$ lower. For Re = 80k case, both clean and rough surface can generate tonal noise. For the clean geometry the main tones have frequency $St = 3.24$ and $St = 4.11$. The noise level for the rough case is lower and the peak frequency is shifted away from the clean case. More tones can be observed at higher frequencies, i.e. $St= 3.25$, $3.98$ and $4.16$. For the Re = 100k cases, very strong tonal noise can be observed at $St = 3.53$, $4.02$, $6.96$ and $8.02$ for the clean geometry, while the most of tones are eliminated from the spectrum with the presence of roughness elements, with the exception of tones with  $St = 5.56$ and $6.15$ with much lower amplitudes. 

However, numerical and experimental results are not directly comparable  due to the number of facts. First, in the simulation, free-stream turbulence level is assumed to be zero due to the lack of proper free-stream turbulence generation tools. However, this free-stream turbulence can lead to early transition to turbulence which eventually can affect the noise generation. For instance, in the later in the figure \ref{fig:visual_streaks} and in the companion work \citet{Alva2024arxiv}, a visualisation on experiments indicate that flow states on airfoil can be slightly different even if at the similar Reynolds number. Second, in the experimental dataset, it is hard to completely split background noise from the spectrum that we are interested in. Beamforming results help in that matter, but some background noise inevitably remains. Finally, and perhaps more significantly, there is a Mach number difference between experiments and simulations (Mach = 0.3 in numerical simulations and Mach = 0.037 for experiments). In the experimental work of \citet{ProbstingS.2022EoMn}, the authors found that at zero incidence the tonal noise regime in the low-moderate Reynolds number regime to be sensitive to the variation in Mach number. Thus, to be able to compare with experimental results, relationship of dominant tonal frequency and Mach number needs to be investigated. This will require an amount of extra work, especially since low Mach number simulation become very expensive due to CFL constrains. However, it is already encouraging that the same qualitative trends related to the use of roughness elements are observed in the both simulation and experiment.

\begin{figure}
    \centering
    \includegraphics[width = 0.9\linewidth]{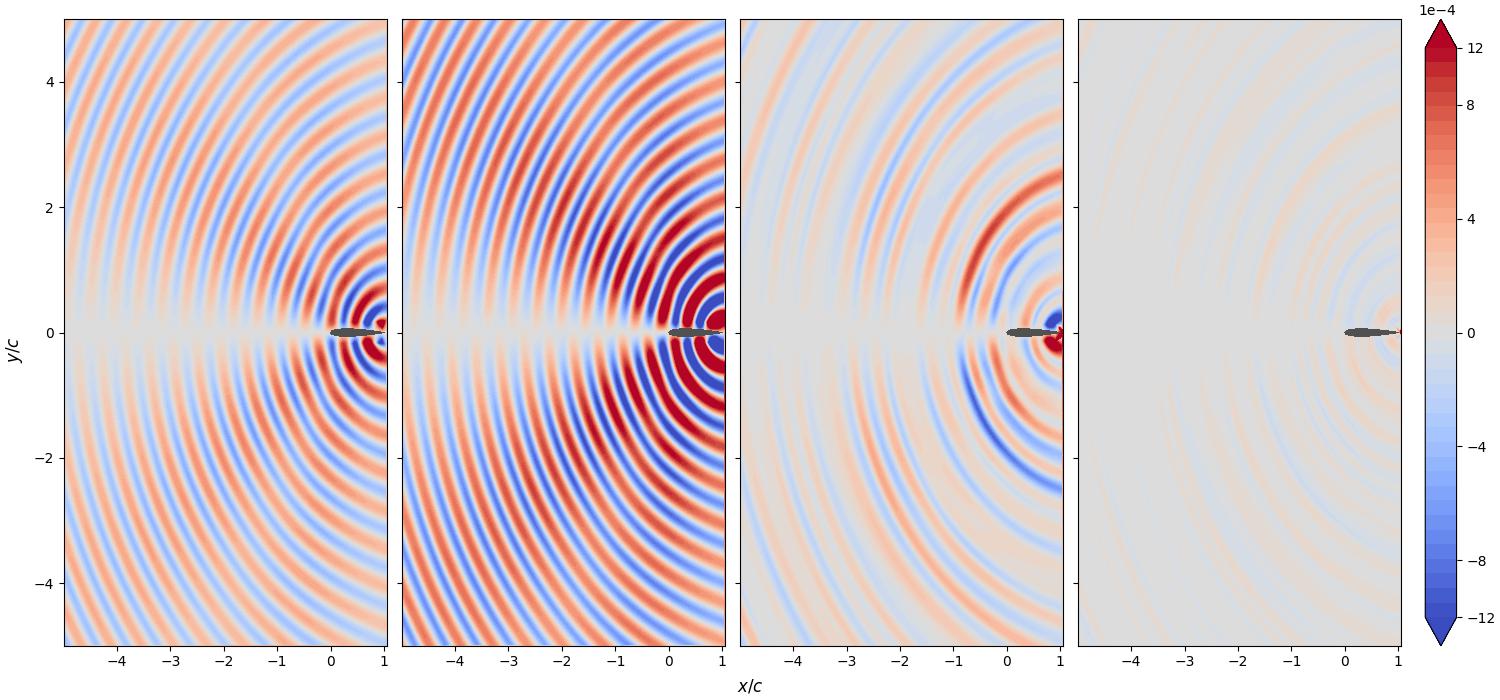}
\caption{Acoustic field contour plot for pressure fluctuation $p'$. The airfoil leading edge point is at $x/c = y/c = 0.0$. From left to right: Re = 80k clean, Re = 100k clean, Re = 80k rough, Re = 100k rough.}
    \label{fig:acous_field}
\end{figure}

Sample snapshots of the two-dimensional acoustic field of all four cases are illustrated in figure \ref{fig:acous_field} with pressure fluctuation $p'$. In the clean cases, strong acoustic radiation can be observed from the trailing edge region. From lower to higher Reynolds number, the amplitude of acoustic waves becomes stronger.  With the presence of roughness elements, amplitude of acoustic waves is lower and wave length is changed compared to the clean cases. At Re = 100k rough case, the acoustic radiation is much lower than other cases and it is hard to identify under the same level range. These observations are consistent with the results of acoustic spectrum in figure \ref{fig:spectrum}.

\subsection{Flow structures}
\begin{figure}
    \centering
    \begin{subfigure}{0.45\textwidth}
        \centering
    \includegraphics[width = \linewidth, trim={0cm 2cm 6cm 10cm},clip]{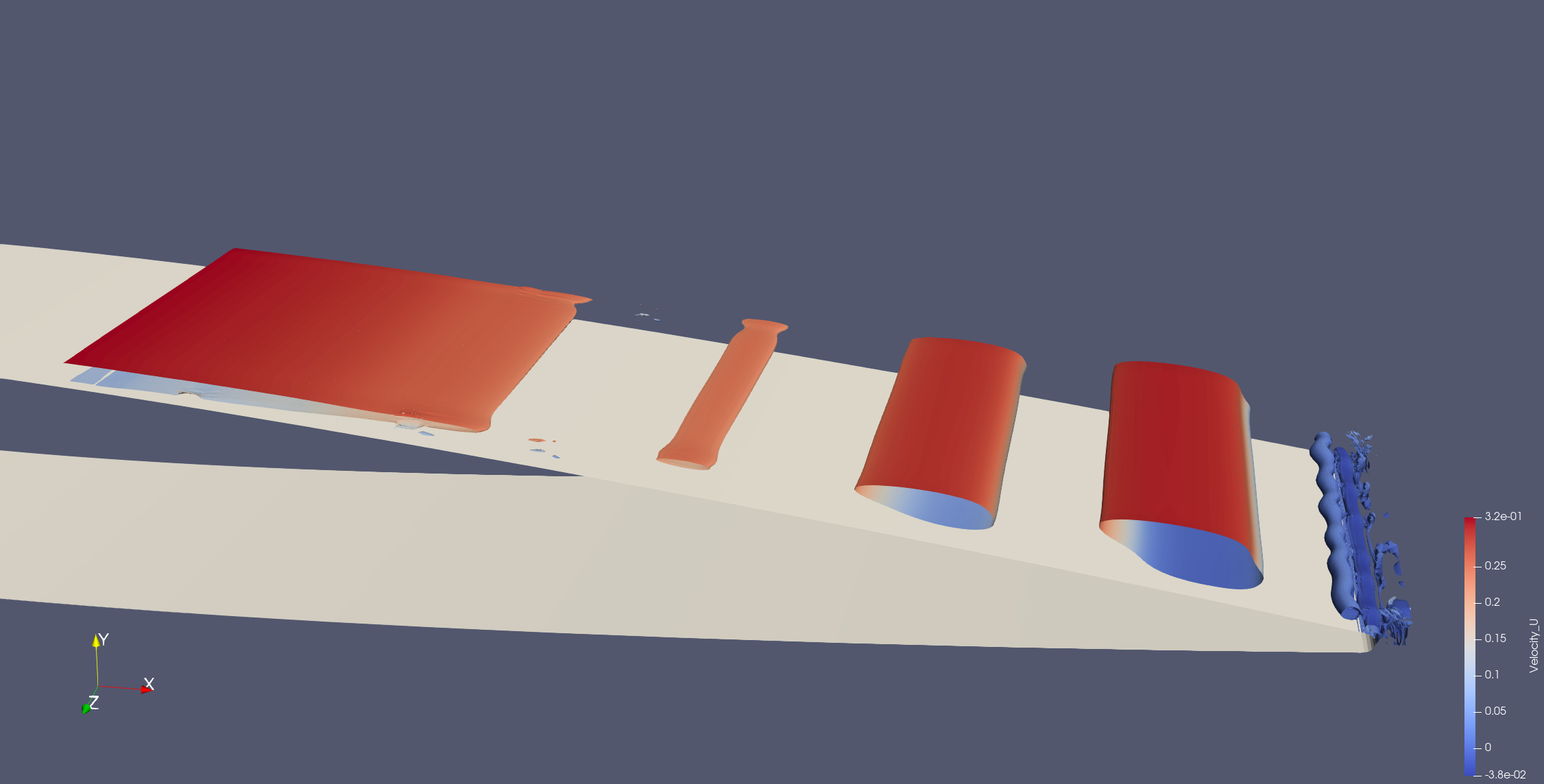}
    \caption{Re = 80k clean}
    \end{subfigure}
    \begin{subfigure}{0.45\textwidth}
        \centering
    \includegraphics[width = \linewidth, trim={0cm 2cm 6cm 10cm},clip]{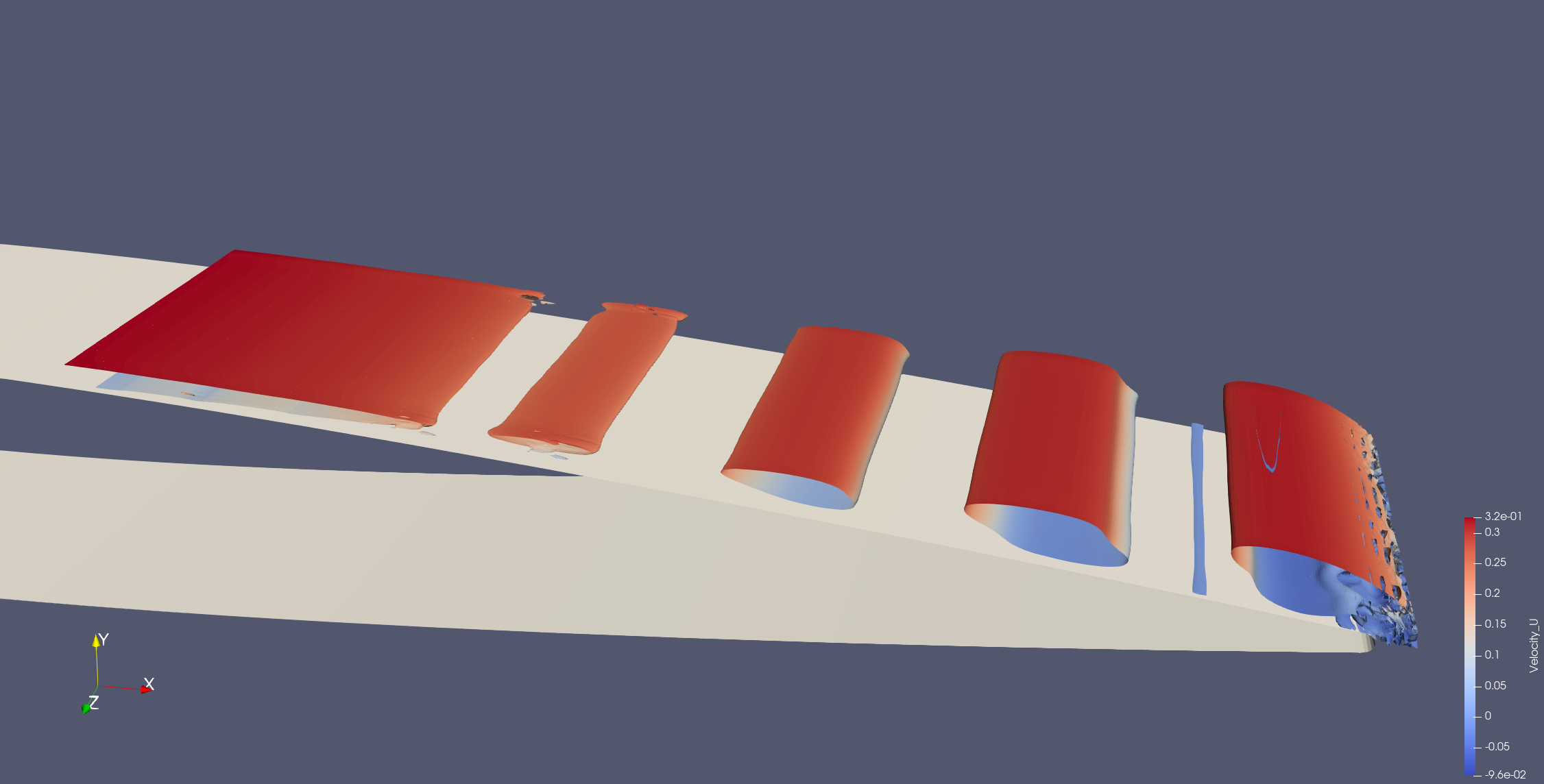}
    \caption{Re = 100k clean}
    \end{subfigure}
    \begin{subfigure}{0.45\textwidth}
        \centering
    \includegraphics[width = \linewidth, trim={0cm 2cm 6cm 10cm},clip]{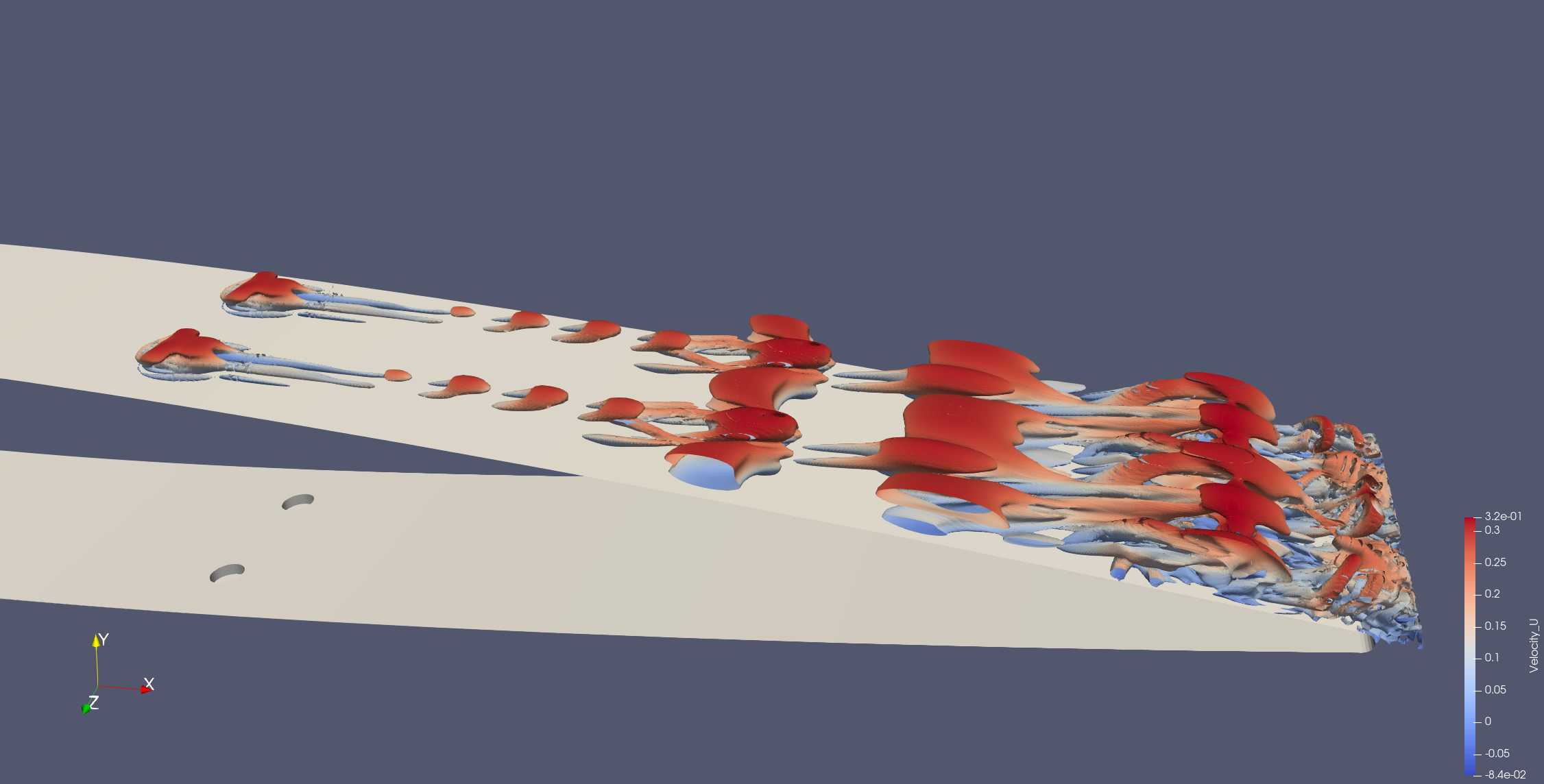}
    \caption{Re = 80k rough}
    \end{subfigure}
    \begin{subfigure}{0.45\textwidth}
        \centering
    \includegraphics[width = \linewidth, trim={0cm 2cm 6cm 10cm},clip]{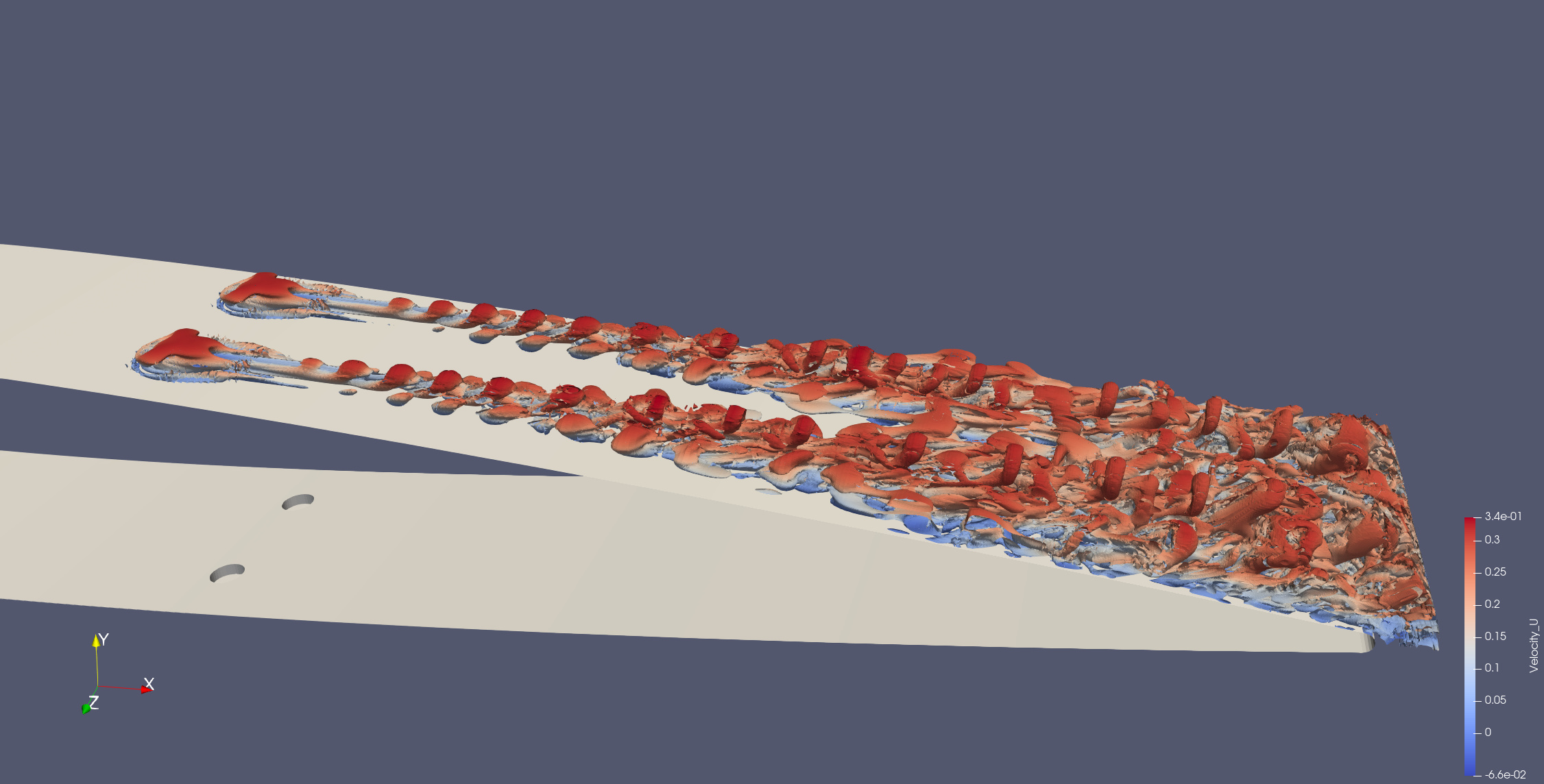}
    \caption{Re = 100k rough}
    \end{subfigure}
\caption{Iso-surface of Q-criterion colored by streamwise velocity $u$.}
    \label{fig:Q}
\end{figure}

As an overview of the flow structures, the visualization of the flow filed is presented in figure \ref{fig:Q} in terms of Q-criterion \citep{Jeong_1995} colored by streamwise velocity component $u$. The Q-criterion is defined as $1/2 ( ||R||^2 - ||S||^2 )$, where $R$ and $S$ are the rotation and strain rate tensors. To enhance visualization and conserve computational resources, a boxed region covering the upper side of the airfoil is utilized for calculating the Q-criterion. Q-criterion can help us to identify the flow states and dominate structures on the airfoil. The Q-criterion value used in the visualizations for clean cases are one order of magnitude higher than those for the rough cases. Videos corresponding to four cases can be seen in the supplementary material.

For clean cases, dominant structures are Kelvin–Helmholtz (K-H) type rollers. These spanwise coherent structures are responsible for tonal noise generation at trailing edge as also identified in the works of \cite{ProbstingS.2015Rotn, RicciardiTulioR.2022Tiap}. With roughness elements on the airfoil surface for both Reynolds numbers, streaky structures are generated by roughness elements. These streaks will develop to lambda structures (also called hair-pins) while propagating downstream and modulating K-H rollers. At the lower Re, K-H type rollers are modulated but maintain certain level of spanwise coherence. An earlier breakdown of K-H rollers to 3D structures is spotted close to the trailing edge. With the increase of Re, the flow structure on the airfoil surface becomes dominated by 3D lambda structures. The trace of spanwise coherent structures is very weak. From these observations, we can infer that these structures may link the acoustic field presented in figure \ref{fig:spectrum} and the hydrodynamic field. Tonal noise is generated by a feedback mechanicsm involving the spanwise coherent K-H rollers. Modulated K-H rollers by streaks which will decrease spanwise coherence and in the end, decrease acoustic amplitude and shift tonal frequencies. Stronger streaks at higher Re could lead to an earlier breakdown to 3D structures which will generate broadband type of noise rather than tonal noise. Further analysis of these aspects will be presented in \S \ref{sec:SPODST}.

\begin{figure}
    \centering
    \begin{subfigure}{0.45\textwidth}
        \centering
    \includegraphics[width = \linewidth]{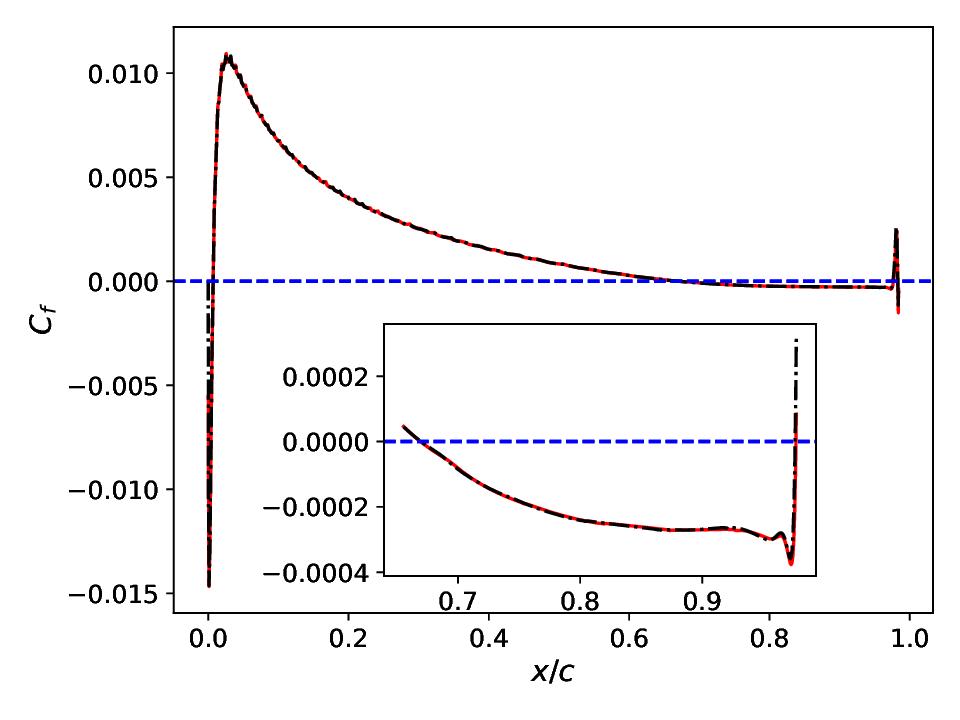}
    \caption{Re = 80k clean}
    \end{subfigure}
    \begin{subfigure}{0.45\textwidth}
        \centering
    \includegraphics[width = \linewidth]{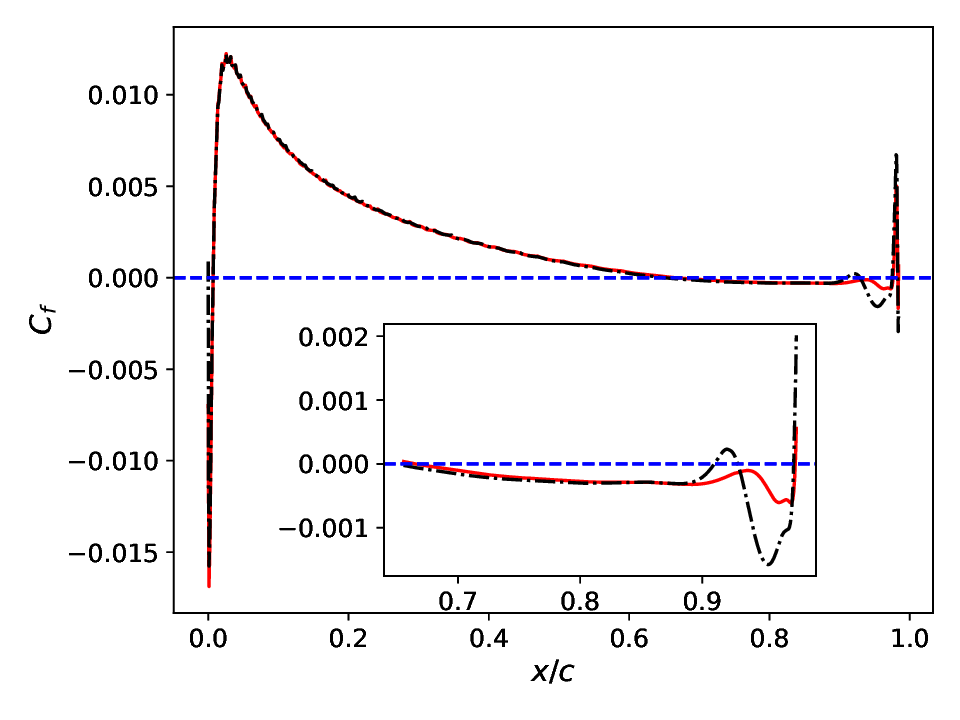}
    \caption{Re = 100k clean}
    \end{subfigure}
\caption{Skin friction coefficient ($C_f$) of span-averaged field. Solid red line and dashed black represent upper and lower sides of the airfoil, respectively. Blue dashed line indicates zero values. }
    \label{fig:cf}
\end{figure}

In the numerical work from \citet{RicciardiTulioR.2022Tiap}, separation bubbles play an important role, as they lead to strong Kelvin-Helmholtz instabilities. In order to reveal the changes of separation bubble induced by streaks, the span- and time-averaged skin friction coefficient ($C_f$) for clean cases are shown in figure \ref{fig:cf}. Here skin friction coefficient is defined as:
\begin{equation}
    C_f = \frac{\tau_w}{0.5\rho_{\infty}U_{\infty}}
\end{equation}
where $\tau_w$ represents the wall shear stress and subscript $\infty$ refers to the free-stream quantities. For the clean cases, a long separation bubble can be spotted on the both sides of airfoil. 
At Re = 80k, the separation bubble initiates at $x/c = 0.65$ and at $x/c = 0.67$ for the higher Re case, extending up to the trailing edge.
A significant difference of $C_f$ at the two sides of airfoil close to the trailing edge can be seen for the higher Re case. This phenomenon has also been documented in the work of \cite{RicciardiTulioR.2020Osta} where a series of 2D simulations were conducted at low Reynolds numbers and Mach numbers. At Re = 100k and Mach = 0.3, the dynamics display symmetry breaking, with a stronger separation bubble at one of the airfoil sides which occasionally switches to the other side. Apparently, our dataset does not cover such a long time evolution, which could be the reason for the  difference of $C_f$ between two sides of the airfoil. Though this could slightly affect the tonal noise characteristics, it will not change the fact that streaks can attenuate tonal noise level. So in order to avoid long and expensive simulations to have fully converged statistics, we keep the current dataset for further analysis.

\begin{figure}
    \centering
    \begin{subfigure}{\textwidth}
    \centering
    \includegraphics[width = 0.9\linewidth,clip]{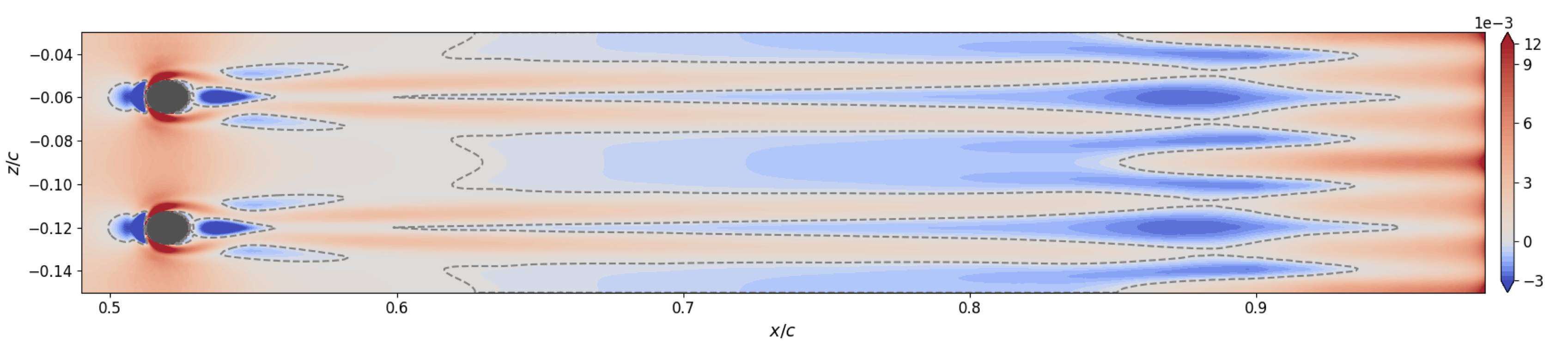}
    \caption{Re = 80k}
    \label{fig:cf_hot_80k}
    \end{subfigure}
    \begin{subfigure}{\textwidth}
    \centering
    \includegraphics[width = 0.9\linewidth,clip]{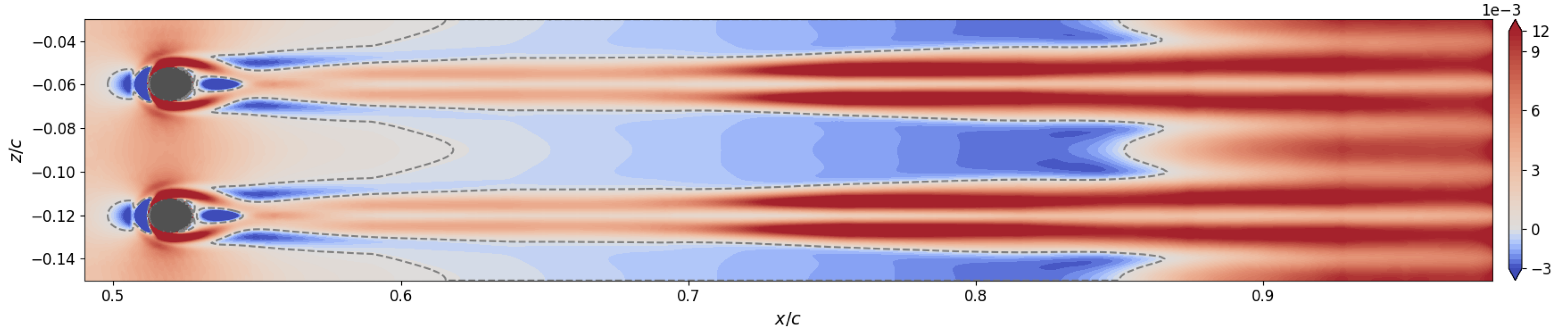}
    \caption{Re = 100k}
    \label{fig:cf_hot_100k}
    \end{subfigure}
    \begin{subfigure}{\textwidth}
    \centering
    \includegraphics[width = 0.8\linewidth,clip]{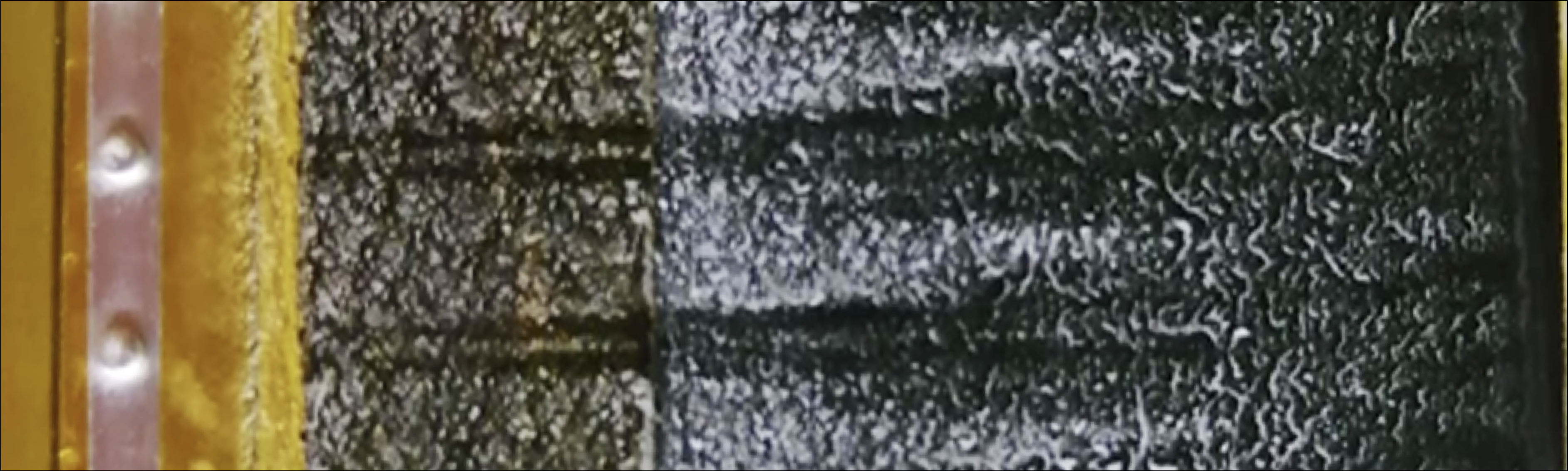}
    \caption{Experimental visualization of streaks}
    \label{fig:visual_streaks}
    \end{subfigure}
\caption{(a, b): $C_f$ hot maps of mean flow field of roughness cases. Red is positive and blue is negative. Dashed grey line is zero-contour line. 
(c): Oil visualization from the experimental campaign at Re = 150,000.}
    \label{fig:cf_hotmap}
\end{figure}

For the cases with roughness elements, the structures on the airfoil are three dimensional. Hence spanwise averaged statistics are not appropriate anymore. Thus, to understand how the existence of structures generated by the roughness elements modifies the surface flow field, time averaged surface skin friction coefficient maps are presented in the figures \ref{fig:cf_hot_80k} and \ref{fig:cf_hot_100k}. The color-map is saturated such that blue indicate negative and red represents positive values. The grey dashed line indicates the zero-contour line.
A re-circulation region can be spotted prior to and after cylinder roughness in the both cases. Downstream of the location of roughness elements, a long separation bubble (characterized by the negative $C_f$ regions) can be spotted in the both cases. However, the separation bubble has been sliced/modulated into sub-domains by the streamwise elongated structures highlighted by the positive $C_f$ values. Fast-speed streaks (indicated by regions of positive $C_f$) are induced downstream of roughness elements. in agreement with earlier studies \citep{FranssonJensH.M.2004Eati, FranssonJensH.M.2005Esot}.
At lower Reynolds numbers, streaks have significantly lower amplitudes. The initial locations of separation bubbles are roughly equivalent to those in the clean case, yet the re-attachment location is further upstream compared to the clean case.
For the high Re case, streaks have higher amplitude right after the roughness elements, maintaining it until the trailing-edge region. The end of the separation bubble moves upstream, to around $x/c = 0.86$. 

Figure \ref{fig:visual_streaks} illustrates flow visualization from the experiment at Re = 150,000 in which the surface oil flow visualization technique is employed to reveal flow patterns, with technical details provided in the first section of this paper. It should be noted that a higher Reynolds number was chosen to present to enhance the visualization effect, although the overall mechanism remains the same. This visualization method captures the fast and slow-speed streaks directly behind roughness elements, with streak traces extending up to around $x/c = 0.9$. This finding is highly consistent with the $C_f$ distribution from the simulation depicted in the same figure.

\section{Spectral and stability analysis}
\label{sec:SPODST}
The observations mentioned above confirm the presence and significance of streaks in modulating spanwise coherent structures, ultimately leading to a reduction in tonal noise. In this section, we will use data driven spectral and local stability analysis methods to quantify the contribution of streaks to reduction of the growth of spanwise coherent structures. In order to accomplish this, a series of cross planes at different streamwise locations ($x/c = 0.55$, $0.65$, $0.75$, $0.85$, $0.95$) are used for analysis. Local coordinates  used here are $z-z'$ and $y-y_w$ where $z'$ represents the center of cylinder roughness and $y_w$ is the wall location. Results with respect to only one period of cylindrical roughness are presented in the following context.

\begin{figure}
    \centering
    \begin{subfigure}{0.45\textwidth}
    \centering
    \includegraphics[width = \linewidth]{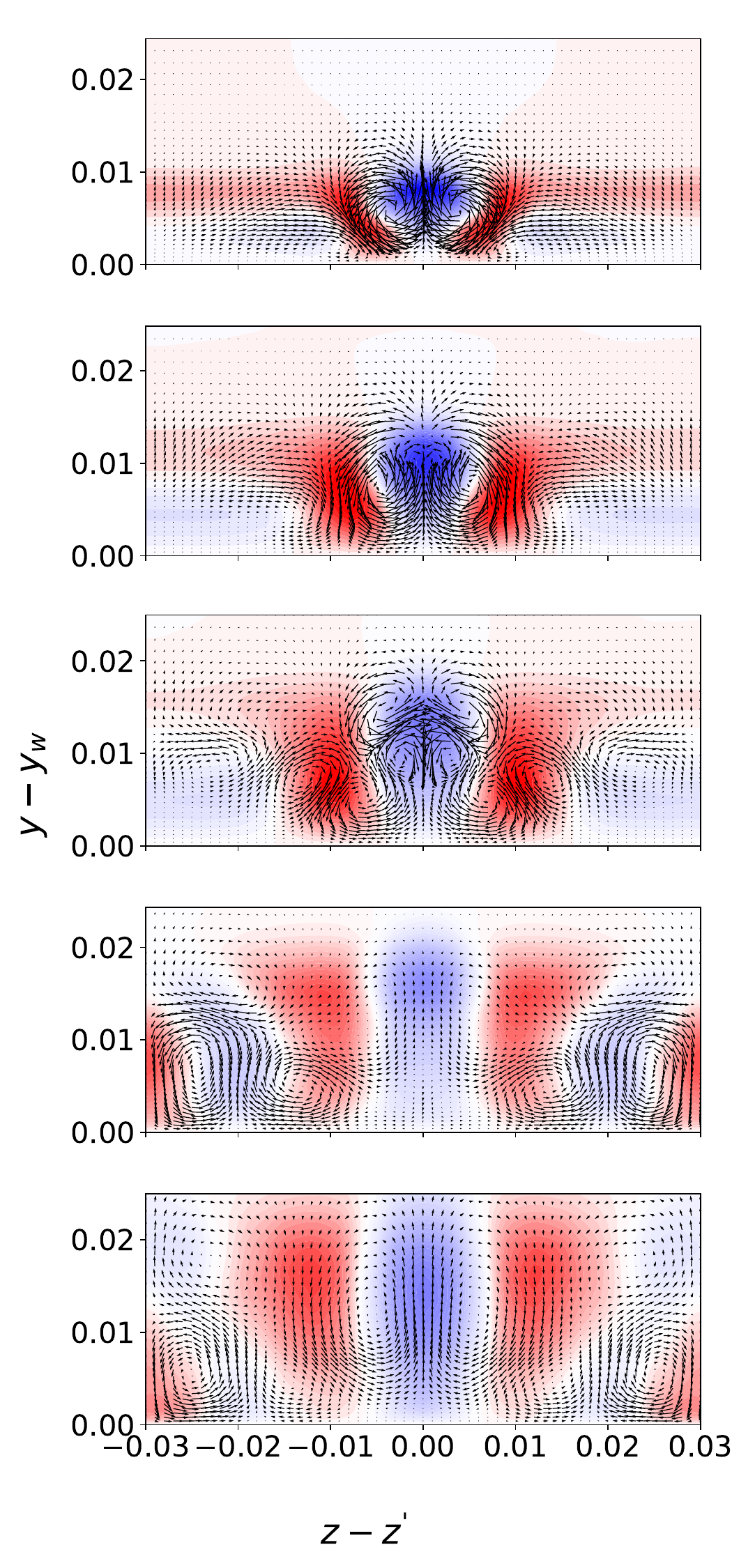}
    \caption{Re = 80k}
    \end{subfigure}
    \begin{subfigure}{0.45\textwidth}
    \centering
    \includegraphics[width = \linewidth]{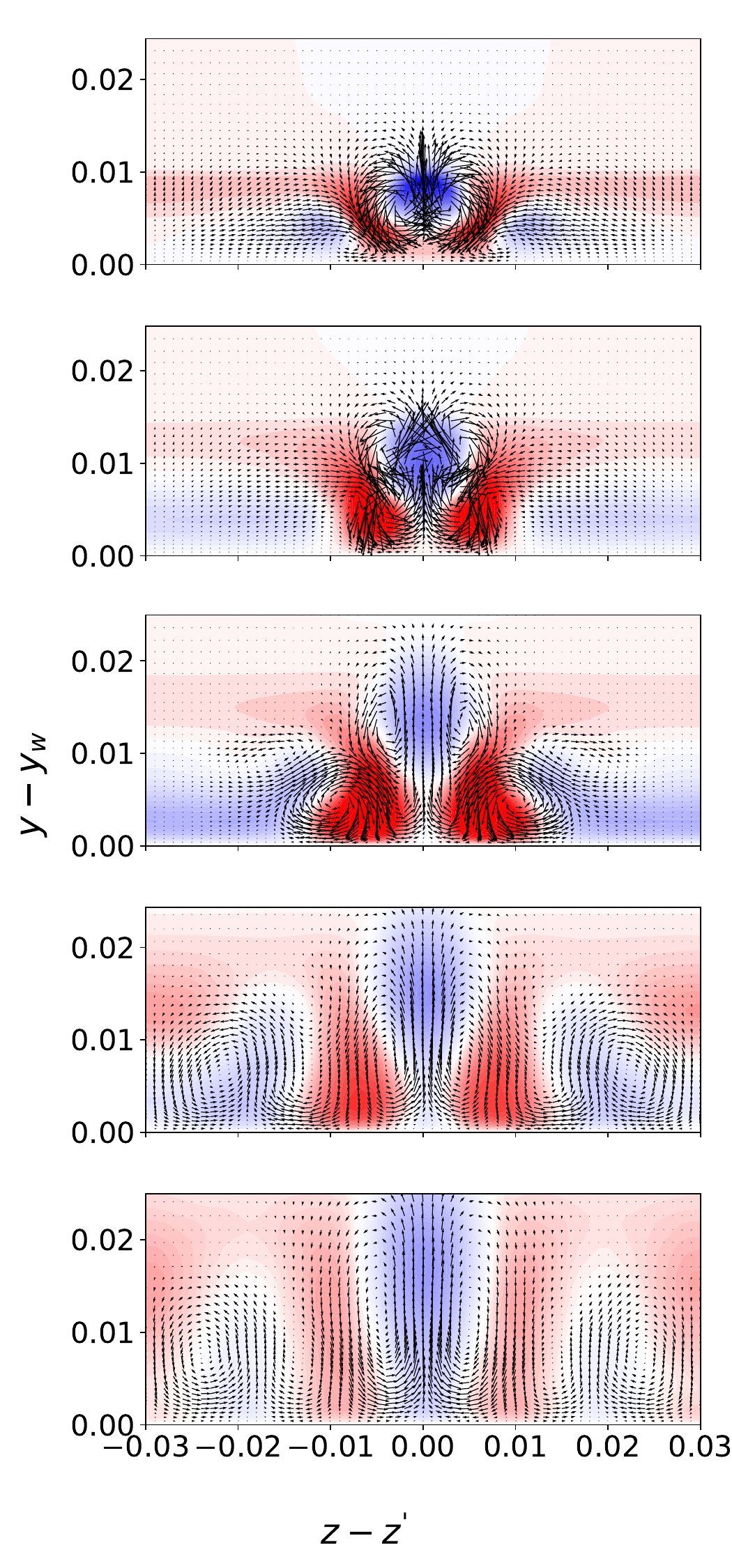}
    \caption{Re = 100k}
    \end{subfigure}
\caption{Contour plots show the wall tangent mean flow $u_t$ where $u_t$ indicates the difference between time averaged and time-spanwise averaged wall tangent mean flow. Arrow fields show time averaged velocity components $v$ and $w$ in the wall normal and spanwise direction respectively.
From top to bottom row, sub-figures show streamwise stations $x/c = 0.55$, $0.65$, $0.75$, $0.85$, $0.95$. }
    \label{fig:station_mean}
\end{figure}

The wall tangential mean flow profile $u_t$ at these locations is shown in figure \ref{fig:station_mean}. Here $u_t$ is defined as the difference between the time-averaged mean profile and the time-spanwise mean profile as an indication of the modification of the mean flow by streaks. In addition, arrow fields generated by the time-averaged wall normal and spanwise velocity components ($v$, $w$) are added to the contour plots to help identify streamwise vortices. The amplitude of the contours and arrows is kept the same across all stations for better comparison.

At the station close to the roughness elements ($x/c = 0.55$), arrow field plot clearly reveals a pair of counter-rotating vortices that transport high-velocity fluid from the free-stream downwards to near wall region and low-velocity fluid from near wall region to high momentum fluid region. This phenomenon is commonly referred to as the ``lift-up" effect mechanism, as detailed by \cite{EllingsenT.1975Solf, LandahlM.T.1980Anoa} and more recently in the review of \cite{BrandtLuca2014TleT}.
Through the lift-up mechanism, a low velocity streak appear further from the wall, along the centreline of the roughness element and high speed streaks appear closer to the wall. As one moves towards the downstream stations, we see the emergence of counter-rotating streamwise vortices which are responsible for the central low-speed streak that develops downstream.  Furthermore, streamwise vortices start to spread in the spanwise direction and their amplitude decreases. For the higher Re case, the overall mechanisms are the same, but more pronounced streaks are present, leading to stronger evidence of fluid transport. This can be seen at $x/c = 0.75$, where the Re = 100k case shows that high momentum fluid occurs much closer to the wall region compared to the lower Re case. Through the lift-up mechanism, the reverse flow is counteracted by high-speed fluid as showed in the Figure \ref{fig:cf_hotmap}.

\subsection{Spectral proper orthogonal decomposition (SPOD)}
The previous visualizations are of steady structures induced by the roughness elements. We now characterise fluctuating fields, which are expected to be directly related to acoustic radiation, using modal analysis. Proper orthogonal decomposition (POD) is a data-driven method which extracts a set of orthogonal basis functions from flow realizations \citep{lumey_stochastic_1970, berkooz_proper_1993} that maximize the mean square energy, considering an appropriate inner product. In the spectral version of POD (termed SPOD \citep{PICARD1999443, towne_schmidt_colonius_2018}), this basis is defined by the eigenvalue decomposition of the cross spectral density (CSD) matrix of the Fourier-transformed realizations at each frequency, ensuring spatial-temporal coherence of the structures. We consider the standard Reynolds decomposition of the flow variables, i.e. $q = \Bar{q}+q'$ where $\bar{q} = \left[\overline{\rho},\overline{u},\overline{v},\overline{w},\overline{T}\right]^\text{T}$ stands for time average and $q'$ is the fluctuation component. Here in this work, we define fluid component as $q' = \left[\rho',u',v',w',T'\right]^\text{T}$ as the density, tangential, wall-normal, spanwise velocity components and temperature fluctuations, respectively. Since SPOD is applied to compressible flow components at the streamwise cuts locations, the norm used here is the compressible energy \citep{ChuBoa-Teh1965Otet, Mack1984, HanifiArdeshir1996Tgic}. 
Following the so-called snapshot method \citep{SchmidtOliverT2020GtSP}, firstly a windowed fast Fourier transform is performed on the fluid field data in time to obtain the flow field $\hat{q}$ (where $\string^$ denotes Fourier-transformed quantities) at each discrete frequency. The decomposition is performed at each frequency based on the inner product between different flow realizations $\hat{q}_{\text{i}}$, given by
\begin{equation}
	\langle \hat{q}_{\text{i}}, \hat{q}_{\text{j}}\rangle = \int_{\Omega}\hat{q}_{\text{j}}^H\mathcal{W}\hat{q}_{\text{i}} d\mathbf{x}= \hat{q}_{\text{j}}^H\textbf{W}\hat{q}_{\text{i}}\text{,}
    \label{eq:inner_product}
\end{equation}
where $\Omega$ is the region of interest, the superscript $H$ denotes the Hermitian transpose, and the discretized weighting operator, $\textbf{W}$, is chosen as such that the sum of the eigenvalues defines a compressible energy (related to the variance of the CSD). Following the original derivation from \citet{ChuBoa-Teh1965Otet}, $\textbf{W}$ is a positive definite weighting tensor:

\begin{equation}
	\textbf{W} =\int_{\Omega}\begin{bmatrix}
    \frac{\overline{T}}{\gamma\overline{\rho}M^2} &  & & &\\
    & \overline{\rho} &  & &\\
    & &\overline{\rho}   & &\\
    & &  &\overline{\rho}  &\\
    & &  &  & \frac{\overline{\rho}}{\gamma(\gamma-1)\overline{T}M^2}
    \end{bmatrix}d\mathbf{x} \text{.}
    \label{eq:energy_weight_SPOD}
\end{equation}
where $\gamma$ is the specific heat ratio and $M$ is the Mach number.

In all the analyses studied here we have used Hamming windows for Welch’s method, since they are commonly used in narrowband applications. The analysis is performed with the numerical code from \citet{SchmidtOliverT2020GtSP}, using a short time fast Fourier transform (FFT) considering $N_{\text{FFT}}=128$ frequency bins per block with an overlap of $75\%$, resulting in the number of blocks $N_{\text{blocks}}=172$ and $\Delta T = 0.03c/U_{\infty}$. In order to increase the resolution of SPOD analysis, data from the two periods of roughness elements on one side of airfoil is treated as two independent realizations.   
As the case with roughness elements does not allow a Fourier transform along the spanwise direction, SPOD modes have a non-trivial z-dependency. To limit the data requirements, we apply SPOD separately at the streamwise positions illustrated in the figure \ref{fig:station_mean}.

\begin{figure}
    \centering
    \includegraphics[width = 0.45\linewidth]{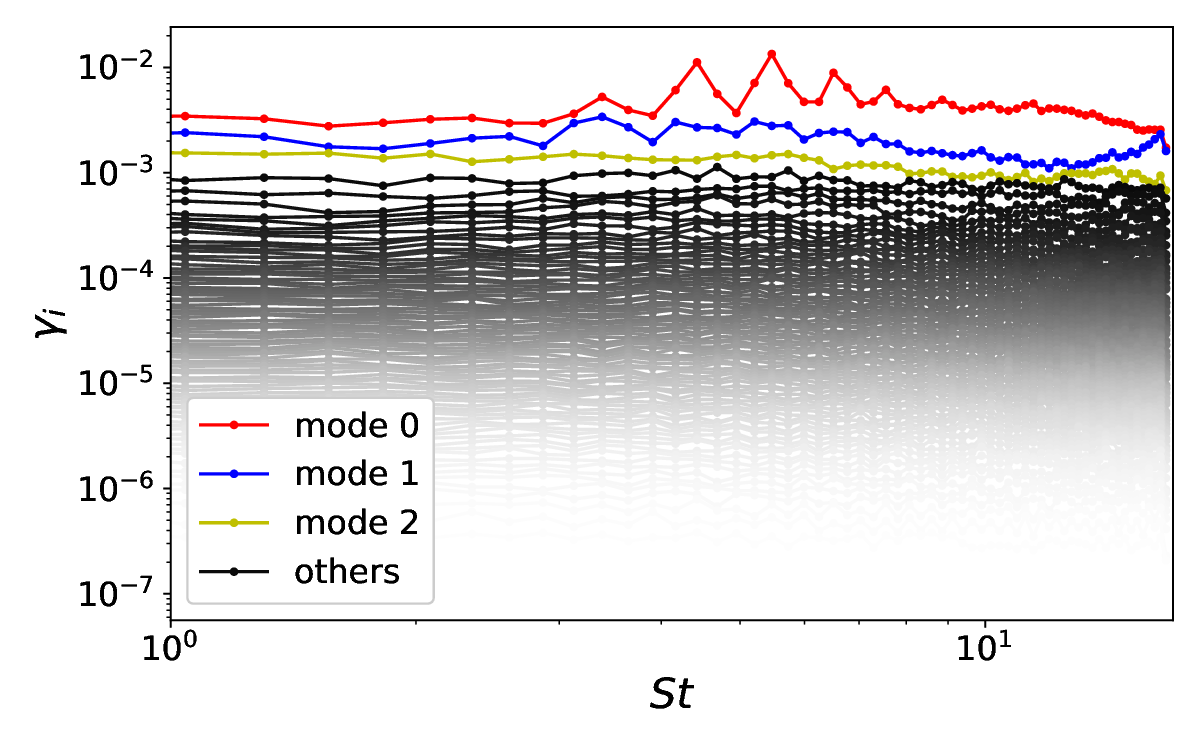}
    \includegraphics[width = 0.45\linewidth]{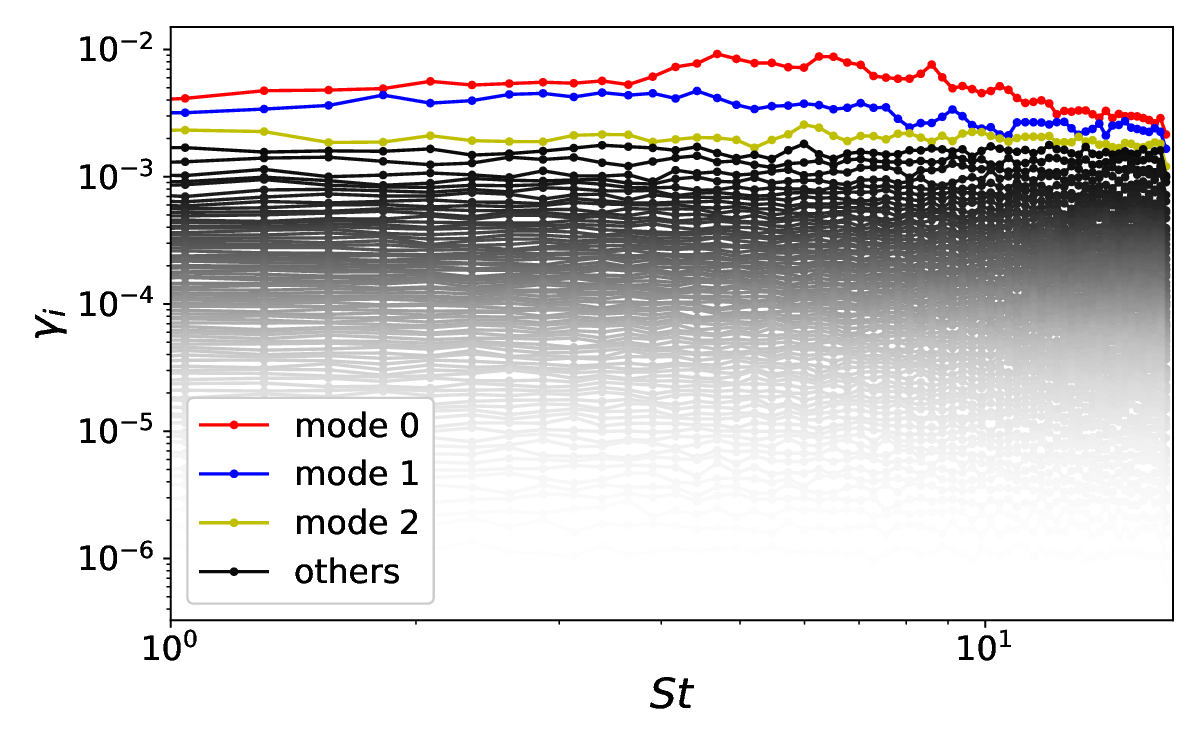}
    \caption{SPOD energy calculated at the station $x/c = 0.75$ at Re = 80k (left) and Re = 100k (right). Data corresponds to the rough cases.}
    \label{fig:spod_energy}
\end{figure}

The energy spectrum ($\gamma_i$) calculated at the station $x/c = 0.75$ is presented in the figure \ref{fig:spod_energy} for both Reynolds numbers. At lower Re, a group of noticeable peaks can be observed in the spectrum at $St = 5.47$, $4.45$ and $6.55$. Those peaks in spectrum are related to tonal noise illustrated in the figure \ref{fig:spectrum}. As expected, for the higher Re, there are no clear peaks in the spectrum and no dominant frequency can be detected. In order to investigate the effects of roughness elements, we choose the frequency $St = 5.21$ in the Re = 100k which corresponds to the tonal noise in the clean case.  It is worthwhile to note that from the spectrum, the leading mode energy ($\gamma_1$) of chosen frequency is around one order of magnitude higher than that of the second mode  ($\gamma_2$) at the interested frequencies, which indicates that the system dynamics is of low rank, i.e. the flow is dominated by the first mode. 
The leading SPOD mode can thus be used as an effective representation for flow structures.

\begin{figure}
    \centering
    \begin{subfigure}{\textwidth}
    \centering
    \includegraphics[width = 1.0\linewidth]{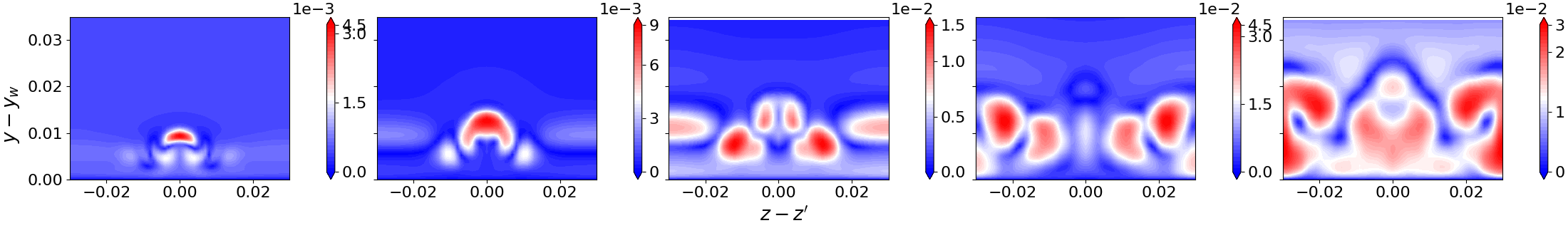}
    \caption{Re = 80k, $St = 5.47$}
    \end{subfigure}
    \begin{subfigure}{\textwidth}
    \centering
    \includegraphics[width = 1.0\linewidth]{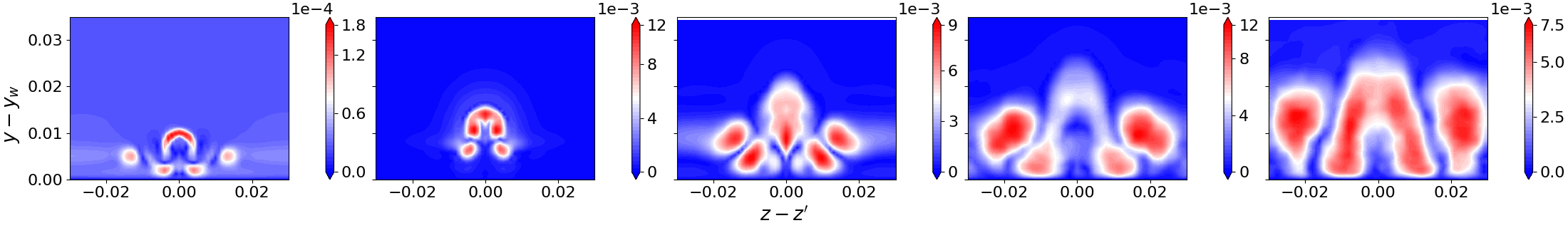}
    \caption{Re = 100k, $St = 5.21$}
    \end{subfigure}
\caption{Absolute value of the leading SPOD mode of streamwise velocity $\Tilde{u}$. From left to right: stations $x/c = 0.55$, $0.65$, $0.75$, $0.85$ and $0.95$.}
    \label{fig:station_all}
\end{figure}

Figure \ref{fig:station_all} shows absolute values of the leading SPOD mode  of wall-tangential velocities $\hat{u}$ at the selected streamwise stations. Note that SPOD modes have unit norm, so to see growth of structures at streamwise stations, SPOD modes are multiplied with square root of its eigenvalue to show actual fluctuation amplitudes in flow. 
From the first and the second station ($x/c = 0.55$ and $0.65$), we see the fast growth of the amplitude of unsteady streak-type structures with the change of almost two order of magnitude for higher Re case. At the station $x/c = 0.75$, for the lower Re case, the growth of spanwise modulated K-H rollers can be identified. Those structures sit at the location $|z-z'|/c \geq 0.015$. For the higher Re, similar patterns are observed. Although the spanwise coherence is lower for Re = 100k, the structures display patterns similar to the Re = 80k case, suggesting that, while of difficult identification in the Q-criterion snapshots, strongly-modulated K-H rollers remain even for higher Reynolds numbers.
From station $x/c = 0.85$ onward ($x/c \geq 0.85$), the amplitude of structures begins to decrease for the higher Re case and continues to increase for the lower Re case. With the growth of the boundary layer, flow structures start to spread out, marking the ongoing process of the transition to turbulence at these stations.
Therefore, in the subsequent discussions, we will primarily focus on the lower Re scenario at around station $x/c = 0.75$.

\begin{figure}
    \centering
    \includegraphics[width = 1.0\linewidth]{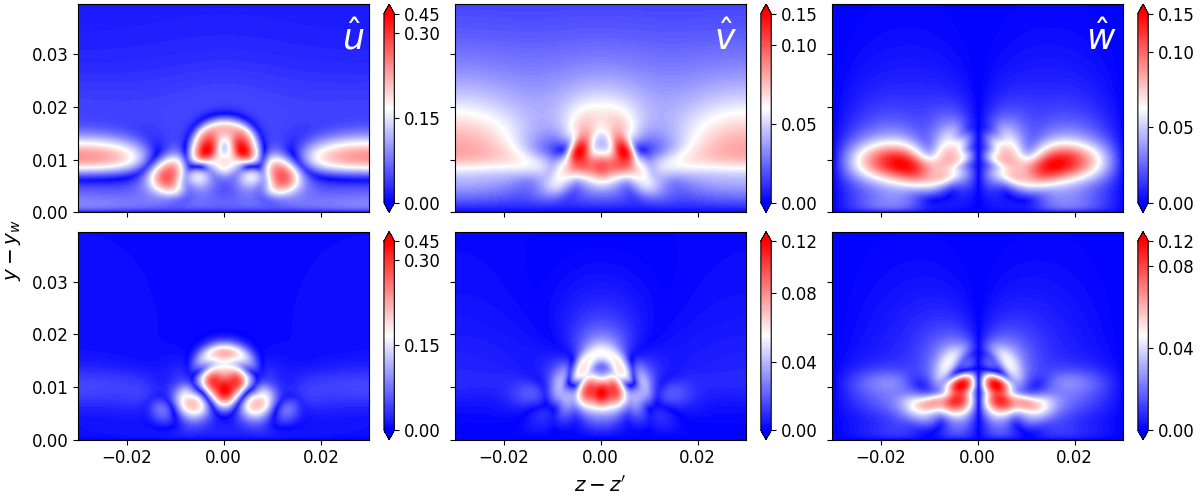}
    \caption{The absolute value of the leading and secondary SPOD modes calculated at the station $x/c = 0.75$ for Re = 80k: the leading SPOD mode (Top row) and the second SPOD mode. From left to right columns: velocity components $\hat{u}$, $\hat{v}$ and $\hat{w}$.}
    \label{fig:spod_modes_80k}
\end{figure}

Figure \ref{fig:spod_modes_80k} depicts the absolute values of the leading and secondary SPOD modes at $x/c = 0.75$ for Re = 80k. At this station, streaks are fully developed and the amplitudes of both streaks and K-H rolls have reached the same order of magnitude. From the figure, the leading SPOD mode describes a K-H modulated by a streak, while the second mode contains only thinner localised structure akin to streak fluctuations. However, at this frequency, the flow field is dominated by the leading SPOD mode according to the energy ranking presented in figure \ref{fig:spod_energy}. Hence, the leading SPOD mode is a good representation of the flow state. The most energetic structures are located $0.01c$ away from the airfoil surface. 

\subsection{Local spatial stability analysis}
In order to investigate how the streaks attenuate growth of the K-H rolls, two-dimensional spatial-local stability analysis in $z-y$ plane is performed at the streamwise station $x/c = 0.75$. 
Here, we are interested in the analysis of hydrodynamic part of the flow field, and because of low Mach number of the flow, incompressible linearized-Navier-Stokes (LNS) formulation is sufficient for the analysis:
\begin{equation}
\begin{gathered}
    \frac{\partial \mathbf{u}'}{\partial t} + (\bar{\mathbf{U}}\cdot \nabla)\mathbf{u}' + (\mathbf{u}'\cdot \nabla)\bar{\mathbf{U}} = -\nabla p' + \frac{1}{Re}\nabla^2\mathbf{u}',  \\
    \nabla\cdot\mathbf{u}' = 0
\end{gathered}
\label{LNS}
\end{equation}
where $\bar{\mathbf{U}}$ indicates the mean flow, upon which linearisation is performed and variables with superscripts $'$ indicate perturbation around the mean flow $\bar{\mathbf{U}}$.  

We carry out a stability analysis in the local setting, neglecting x-variation of the mean flow. Hence, the ansatz for the perturbation has the form:
\begin{equation}
    \{u',v',w',p'\} = \{\hat{u},\hat{v},\hat{w},\hat{p}\}(y,z)e^{i(\alpha x - \omega t)}
    \label{ansatz}
\end{equation}

Substituting the ansatz \ref{ansatz} into the LNS equation \ref{LNS} with assumption of slow variation in the streamwise direction, the following stability problem established:
\begin{equation}
\begin{gathered}
    i\alpha\hat{u} + \frac{\partial\hat{v}}{\partial y} + \frac{\partial\hat{w}}{\partial z}, \\
    -\bar{U}\alpha\hat{u} + \xi\hat{u} + i\hat{v}\frac{\partial\bar{U}}{\partial y} + i\hat{w}\frac{\partial\bar{U}}{\partial z} - \alpha \hat{p} + \alpha^2\frac{i}{Re}\hat{u} = -\omega\hat{u} = 0, \\
    -\bar{U}\alpha\hat{v} + \xi\hat{v} + i\hat{v}\frac{\partial\bar{V}}{\partial y} + i\hat{w}\frac{\partial\bar{V}}{\partial z} - \frac{\hat{p}}{y} + \alpha^2\frac{i}{Re}\hat{v} = -\omega\hat{v}, \\
    -\bar{U}\alpha\hat{w} + \xi\hat{w} + i\hat{v}\frac{\partial\bar{W}}{\partial y} + i\hat{w}\frac{\partial\bar{W}}{\partial z} - \frac{\hat{p}}{z} + \alpha^2\frac{i}{Re}\hat{w} = -\omega\hat{w}, \\
    \xi = i\bar{V}\frac{\partial}{\partial y} + i\bar{W}\frac{\partial}{\partial z} - \frac{i}{Re}(\frac{\partial^2}{\partial y^2} + \frac{\partial^2}{\partial z^2})
\end{gathered}
\end{equation}

Let $\mathbf{q} = \{\hat{u},\hat{v},\hat{w},\hat{p}\}^T$, the equation system can be rearranged to
\begin{equation}
    \left(\alpha\mathcal{A}_1 + \alpha^2\mathcal{A}_2 \right)\mathbf{q} = \left(\omega\mathcal{B} - \mathcal{A}_0 \right)\mathbf{q}
\end{equation}
where $\mathcal{A}_0,\mathcal{A}_1,\mathcal{A}_2$ and $\mathcal{B}$ are functions of the mean flow and flow parameters. Following the spatial stability, we take $\omega \in \mathcal{R}$ to be the temporal frequency input while $\alpha \in \mathcal{C}$ is the complex output eigenvalue. Since we are dealing with hydrodynamic waves whose wavenumbers are lower than one, the $\alpha^2$ terms can be neglected, leading to a linear eigenvalue problem.

\begin{figure}
    \centering
    \includegraphics[width = 0.8\linewidth]{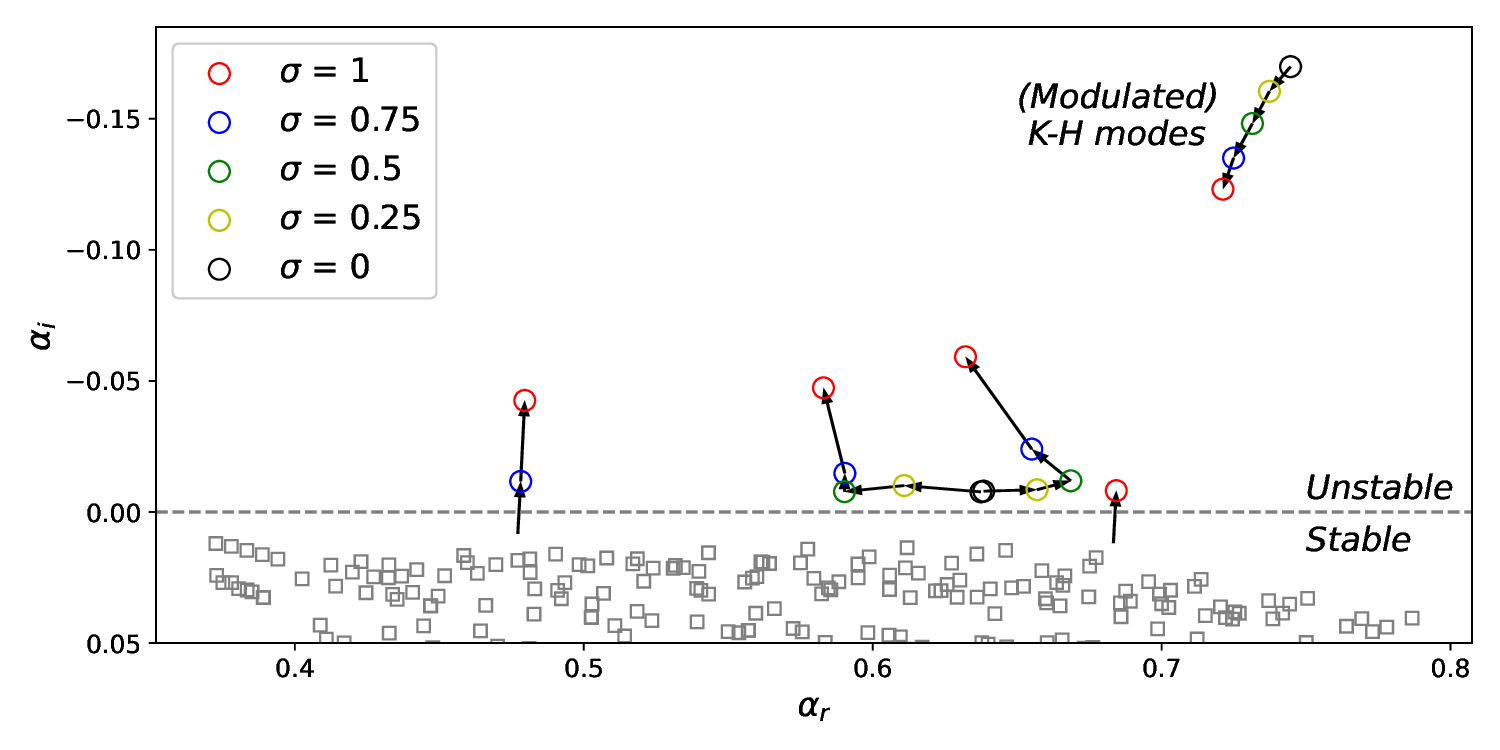}
    \caption{The spectrum traced with gradual increase of mean state modulation parameter $\sigma$. Open circles stand for the unstable modes and open squares are stable modes.}
    \label{fig:track_modes_spectrum}
\end{figure}

\begin{figure}
    \centering
    \includegraphics[width = 1\linewidth]{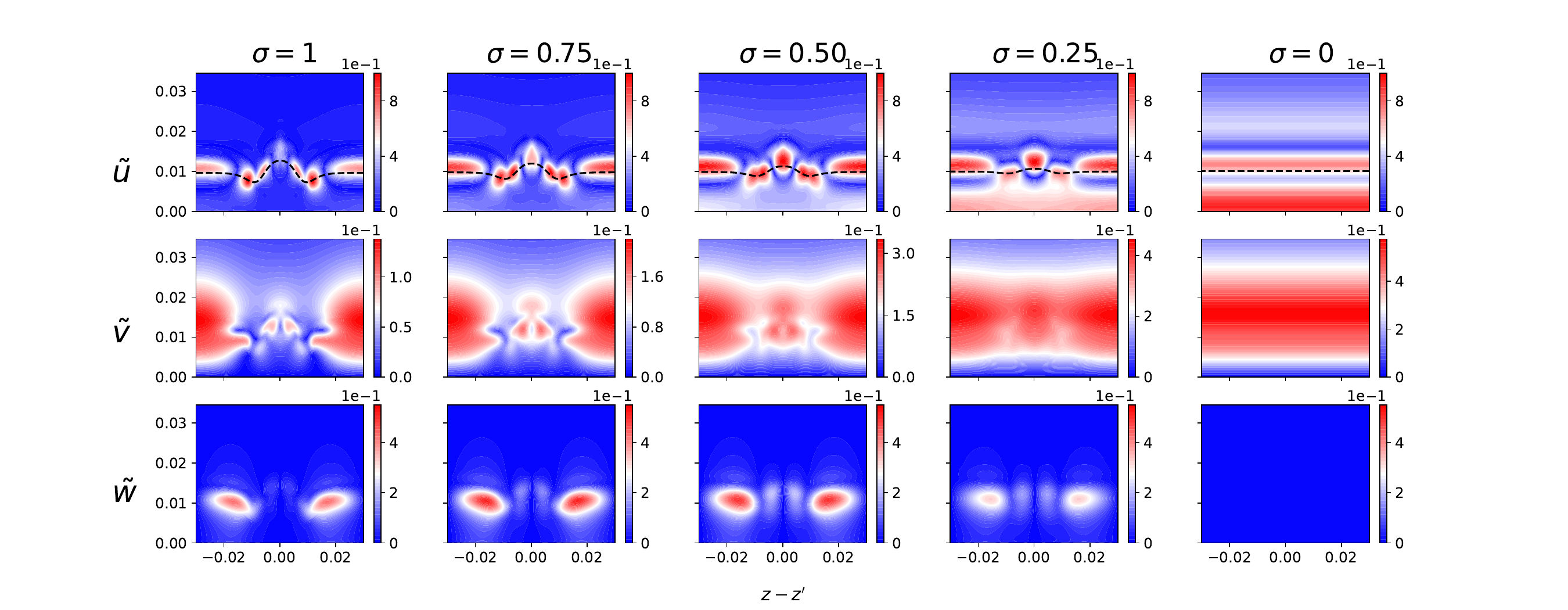}
    \caption{The leading eigen-modes (K-H modes) traced with gradual increase of mean state modulation parameter $\sigma$. $\sigma$ changes from 1 to 0 from left to right columns. Top, middle and bottom row stands for the velocity components $\Tilde{u}$, $\Tilde{v}$ and $\Tilde{w}$ respectively. Black dashed line indicates the critical layer. }
    \label{fig:track_modes}
\end{figure}

The effect of roughness-generated streaks is investigated through tracking the unstable modes from the mean state without roughness elements to the one with roughness elements \citep{LajusF.C.2019Ssao} by a continuous variation of the mean flow ($\mathbf{\bar{U}}$) as:  
\begin{align}
    \mathbf{\bar{U}} = (1 - \sigma) \mathbf{\bar{U}_{clean}} + \sigma \mathbf{\bar{U}_{rough}}
\end{align}
where $0\le \sigma \le 1$. 
The tracking mode technique aids in identifying growth changes between two distinct states. Ideally, continuous change of the parameter $\sigma$ allows us to trace the progression of an unstable mode through these states. For the sake of reducing computational effort, we utilize five discrete values of $\sigma$, which are sufficient to demonstrate the evolution of the eigenvalues.
The spectrum is illustrated in the figure \ref{fig:track_modes_spectrum}. As can be seen there, there are three unstable modes in the clean case. The most unstable mode stands for two-dimensional K-H instability, with the mode shape depicted in the right subplot of figure \ref{fig:track_modes}. The other two modes have identical low growth rate and streamwise wavenumber, corresponding to degenerate oblique K-H modes with positive and negative spanwise wavenumbers, as in \cite{LajusF.C.2019Ssao}. While increasing the value of the parameter $\sigma$, the growth rate of the most unstable mode is decreasing from $\alpha_i = 0.1699$ to $\alpha_i = 0.1231$. However, the other two unstable modes in the clean state diverge and become more unstable with increasing $\sigma$. Two other modes appear from the stable regime, becoming unstable as $\sigma$ approaches 1.  

The shape of the most unstable mode for various values of $\sigma$ are presented in figure \ref{fig:track_modes}. 
Note that eigenvectors are normalised respect to the maximum value of streamwise velocity component $\Tilde{u}$ at each $\sigma$ value. 
From clean state ($\sigma = 0$) to rough state ($\sigma = 1$), a progressive spanwise modulation appears at the middle of domain. The K-H mode is heavily modulated in this region. The black dashed line denotes the critical layer, where $U=\omega/\alpha$, which is the location where the modulated modes peak. The appearance of streaks increases the amplitude of spanwise velocity component $\Tilde{w}$ it while decreases strength of the wall-normal velocity component $\Tilde{v}$. Comparing the eigenmode for $\sigma=1$ with the leading SPOD mode at the same station, the modes show many similarities, indicating that the SPOD mode is indeed representative of a spanwise modulated K-H roller.

\subsection{SPOD vs. Local stability analysis}
From the spatial stability analysis, it can be conclude that streaks can be used to decrease the growth of K-H instabilities. However, we are using mean flow rather than base state of flow due to the difficulty of getting a laminar solution in the present setting. In order to verify this result, one can calculate growth of hydrodynamic waves from the leading SPOD mode. Nonetheless, due to the existence of strong acoustic waves in the field, SPOD modes contain energy of the hydrodynamics and acoustics perturbations. In order to extract hydrodynamic energy from the total energy, we utilize the bi-orthogonality of state and adjoint eigenfunctions, see e.g. \citet{Tumin2011, Nima2016}. In order to use our existing tool, we consider the problem in the framework of temporal stability analysis.
Consider the direct and adjoint problem in a compact form as
\begin{equation}
    \mathbf{N} \mathbf{q} = \omega \mathbf{M} \mathbf{q}
    \label{direct}
\end{equation}
\begin{equation}
    \mathbf{N}^+ \mathbf{q}^+ = \omega^+ \mathbf{M}^+ \mathbf{q}^+
\end{equation}
where superscript $+$ indicates the adjoint and $\omega^+=\text{conjg}(\omega)$.
Let us define the weighted inner product with operator M as the weight operator as
\begin{equation}
    \left<\mathbf{a},\mathbf{b}\right>_\mathbf{M} = \left<\mathbf{M}\mathbf{a},\mathbf{b}\right> = \int_{z_{min}}^{z_{max}}\int_0^{\infty} \mathbf{b}^H\mathbf{M}\mathbf{a}\ dy\ dz,
\end{equation}
where $H$ denotes complex conjugate transpose. Then the bi-orthogonality yields

\begin{equation}
    \left<\mathbf{q}_i^{\ },\mathbf{q}^+_j\right>_\mathbf{M}=c\delta_{ij},
\end{equation}
where $\delta_{ij}$ is the Kronecker delta function. Therefore, by projecting the LES solution onto the adjoint of hydrodynamics modes one extracts their amplitude \citep{RODRIGUEZ2015308}. The amplitude of hydrodynamics wave, $C_{h}$, at each x location can be evaluated through:

\begin{equation}
    C_{h} = \frac{\left<\hat{\mathbf{\Phi}}, \mathbf{q}_{h}^+\right>_\mathbf{M}}{\left<\mathbf{q}_{h}^{\ }, \mathbf{q}_{h}^+\right>_\mathbf{M}}
    \label{bi-ortho}
\end{equation}
Here,  $\mathbf{q}_{h}$ is the eigenvector corresponding to the hydrodynamic mode and $\hat{\mathbf{\Phi}}$ = $\left<\hat{\rho}_s,\hat{u}_s,\hat{v}_s,\hat{w}_s,\hat{T}_s\right>$ is the vector containing the SPOD mode. The denominator in equation \ref{bi-ortho} can be considered as a normalisation constant for adjoint eigenfunction at each station. It is also important to note that the computed hydrodynamic mode amplitudes depends on the normalisation of the direct hydrodynamic eigenmode. Here, we have normalised hydrodynamic eigenmode, $\mathbf{q}_{h}$, with maximum streamwise velocity.

\begin{figure}
    \centering
    \includegraphics[width = 0.8\linewidth]{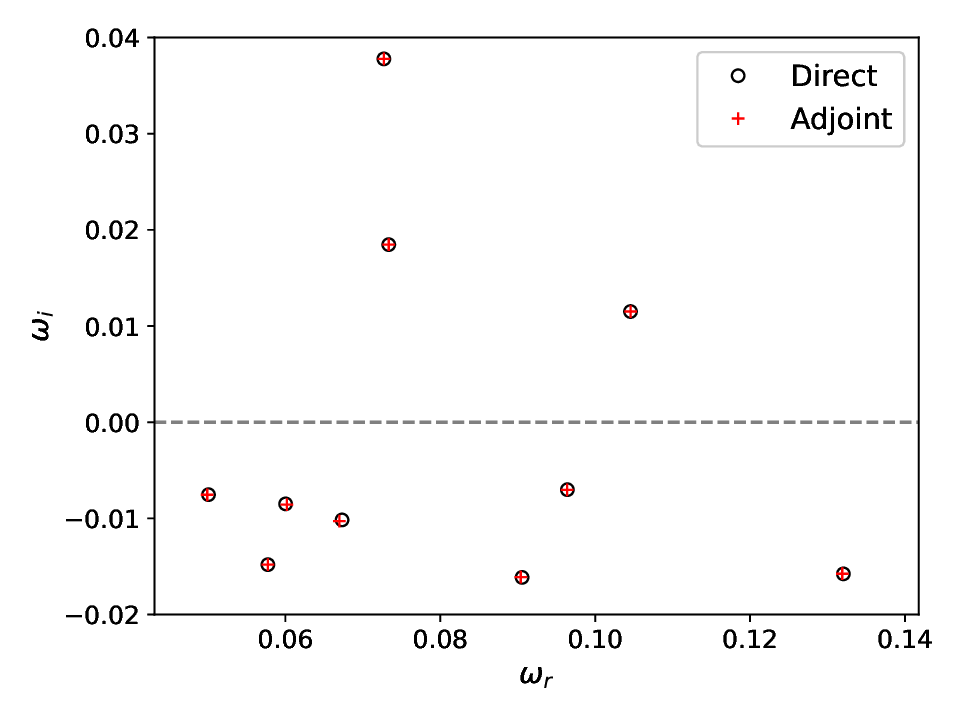}
    \caption{Eigenvalues of direct and adjoint modes calculated at the station $x/c = 0.75$. Note that adjoint eigenvalues are complex-conjugated.}
    \label{fig:stab_eig_conj}
\end{figure}

\begin{figure}
    \centering
    \includegraphics[width = 1.0\linewidth]{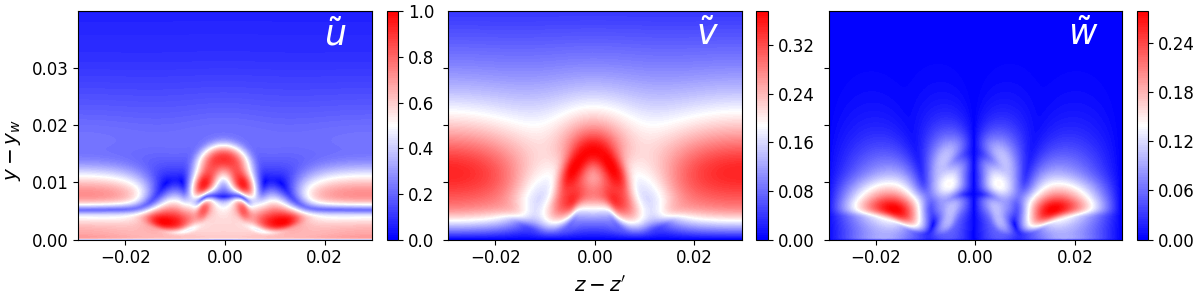}
    \includegraphics[width = 1.0\linewidth]{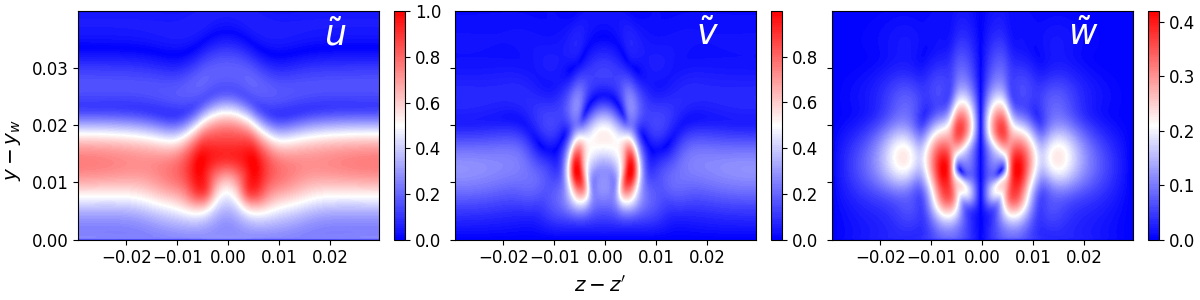}
    \caption{The absolute value of the most unstable mode calculated at the station $x/c = 0.75$. From left to right columns: velocity components $\Tilde{u}$, $\Tilde{v}$ and $\Tilde{w}$. Upper row: direct mode; lower row: adjoint mode.}
    \label{fig:stab_mode_abs}
\end{figure}

The eigenvalues of direct and adjoint analysis are illustrated in the figure \ref{fig:stab_eig_conj} at $x/c = 0.75$. The complex-conjugated adjoint eigenvalues overlap with the direct eigenvalues which indicates the convergence of mesh and resolution.
The most unstable hydrodynamic mode is shown in figure \ref{fig:stab_mode_abs}. The direct mode is closely aligned with the spatial and SPOD analyses. Furthermore, the adjoint mode reveals the sensitivity of flow, and the least unstable mode indicates sensitivity of the spanwise-modulated K-H instability.

Five close stations $x/c = 0.75, 0.75 \pm 4\times10^{-4}, 0.75 \pm 8\times10^{-4}$ are used to calculate the growth of K-H modes around the station $x/c = 0.75$. For this purpose, first, the amplitude of the dominant K-H mode is extracted at theses stations using the projection given in equation (\ref{bi-ortho}). Then the maximum absolute value of spanwise-averaged velocity field (to extract the two-dimensional K-H mode) is used to calculate the growth rate as
\begin{equation}
    \text{growth rate}=\frac{1}{|u|_{max}}\frac{d |u|_{max}}{d x}.
\end{equation}

The results are listed in table \ref{tab:growthrate}. As can be seen there, the projected growth rates of the K-H mode are very close to the prediction given by the spatial stability analysis. This confirms the proposed mechanism for reduction of tonal noise: the introduction of roughness elements induces steady streaks, which render the laminar separation bubble three-dimensional, creating a spanwise modulation of the Kelvin-Helmholtz instability of the shear layer, with a significant reduction of the growth rate. This weakens one of the components of the feedback loop leading to tonal airfoil noise, reducing tones emitted to the far field.  

\begin{table}
    \centering
    \begin{tabular}{c|c c c c }
        Re = 80k & (Clean) Stability  & (Clean) SPOD & (Rough) Stability & (Rough) SPOD\\
        \hline 
        Growth rate & 0.1699 & 0.1613 & 0.1231 & 0.1201
    \end{tabular}
    \caption{Growth rate at station $x/c = 0.75$ from spatial stability analysis and projected SPOD modes via equation \ref{bi-ortho} for both clean ($\sigma = 0$) and rough ($\sigma = 1$) states.}
    \label{tab:growthrate}
\end{table}

\section{Conclusion}
\label{sec:conclusions}
In this article, we report the numerical results of investigations of tonal trailing-edge noise of a NACA0012 and its control. We have used surface roughness elements to attenuate amplitudes of the far-field acoustics. 
We have performed a series of wall-resolved large eddy simulations with the spectral-element solver PyFR with and without surface roughness elements at two Reynolds numbers (Re=80,000 and 100,000). 
The acoustic spectrum illustrate trends similar to those observed in the experiments in our companion work \citep{Alva2024arxiv}. For the clean geometry, the acoustic spectrum is dominated by the main tone ($St = 5.47$ for Re = 80k and $St = 5.21$ for Re = 100k). With the presence of roughness elements, in the lower Re case, the main tone splits into multiple tones spreading to a wider frequency regime. Those tones have lower amplitude than the main one in the clean case. For higher Re, tonal noise is eliminated from the spectrum and a broadband spectrum is identified.

Q-criterion iso-surface contour plots indicate that two dimensional spanwise coherent structures (Kelvin–Helmholtz (K-H) rollers) are the dominant flow structures over the clean airfoil surface and the main source of radiated acoustics, as shown in literature \citep{ProbstingS.2015Rotn, RicciardiTulioR.2022Tiap}. 
The introduction of roughness elements leads to the formation of streamwise elongated structures, which further downstream can be identified as streaks. These streaks modulate Kelvin-Helmholtz rolls, reducing their spanwise coherence. At higher Re, an early transition of these structures into three-dimensional forms occurs. Tonal noise is two dimensional and according to the scattering condition \citep{Nogueira2017Ampf}, lower spanwise coherence is expected to decrease the amplitude of the tonal noise.
Furthermore, the skin friction coefficient $C_f$ illustrates that a long separation bubble exists from around 65\% of the chord till the trailing edge. The presence of roughness element modulates and slices this separation bubble into subdomains and pushes the re-attachment point further upstream. The amplified and modulated K-H roller breaks down earlier on the airfoil surface which generated structures with low spanwise coherence; this weakens one of the components of the feedback loop leading to tonal noise. For the higher Re, the streaks are even stronger which results even lower spanwise coherence close to the trailing edge.

Spectral proper orthogonal decomposition (SPOD) has been employed to analysis dominant flow structures at a series of streamwise stations. At the lower Re, the leading SPOD mode is consistent with a K-H instability modulated in the spanwise by the presence of streaks.
Two dimensional spatial stability analysis is performed at the streamwise station $x/c = 0.75$ where the spanwise-modulated K-H mode is well developed according to the SPOD analysis. Spatial stability analysis is performed on weighted average of mean flow states between clean geometry and rough geometry controlled by the shape factor $\sigma$. By assigning varies $\sigma$ values, we can track unstable modes between two states. Results illustrate that the existence of streak attenuate the growth of K-H instability from 0.1699 to 0.1231. This reduction of growth rate provides theoretical evidence of the mechanism reducing tonal noise radiation.  This theoretical result is confirmed via calculating the growth rate from SPOD modes. The SPOD modes contain a large portion of energy associated with the acoustic filed. Hence, to separate hydrodynamic energy from acoustic energy, we employed a bi-orthogonality relation to extract the amplitude of the K-H mode from the leading SPOD mode. The result showed that the growth of K-H instability from clean geometry to rough geometry decreases from 0.1613 to 0.1201 which is very close to the prediction by the stability analysis.

Overall, the analysis of the simulations presented in this work contributes to a better understanding of the mechanisms behind tone attenuation and suppression, as discussed in the experiments of our companion paper \citep{Alva2024arxiv}. The key aspects are related to the formation of streaks that render the laminar separation bubble three-dimensional, resulting in  a spanwise modulation of Kelvin-Helmholtz modes and a reduction of their growth rates. These effects are expected to weaken the feedback loop leading to tonal noise radiation. Future research can build on these theoretical insights to develop more refined designs of roughness elements aiming at the attenuation of the  undesirable tonal noise from airfoils.

\section*{Acknowledgement}
Zhenyang Yuan would like to acknowledge Swedish Research Council for supporting current work under Grant 2020-04084. The computations were performed on resources by the National Academic Infrastructure for Super-computing in Sweden (NAISS) at the LUMI super computer cluster in Finland and Dardel super computer at PDC KTH, Sweden.
El\'ias Alva would like to acknowledge CAPES, under the Brazilian Ministry of Education (MEC) for supporting experimental campaigns and  FAPESP project for providing funding for equipment.

\bibliographystyle{jfm}
\bibliography{jfm}

\end{document}